\documentstyle[12pt]{article}
\newcommand\Journal[4]{{#1} {\bf #2}, #3 (#4)}

\newcommand\NPB{{\em Nucl. Phys.} B}
\newcommand\PLB{{\em Phys. Lett.}  B}
\newcommand\PRL{\em Phys. Rev. Lett.}
\newcommand\PRD{{\em Phys. Rev.} D}

\newcommand\be{\begin{equation}}
\newcommand\ee{\end{equation}}
\newcommand\beq{\be}
\newcommand\eeq{\ee}
\newcommand\bea{\begin{eqnarray}}
\newcommand\eea{\end{eqnarray}}

\newcommand{\msq}{{m_{\tilde{q}}}}
\newcommand{\mq}{{m_{q}}}
\newcommand{\mgluino}{{m_{\tilde{g}}}}
\newcommand{\quark}{{\it q}}
\newcommand{\squark}{{\it \~{q}}}
\newcommand{\gluino}{{\it \~{g}}}
\newcommand{\gluon}{{\it g}}
\newcommand{\PSbox}[3]{\mbox{\rule{0in}{#3}\includegraphics{#1}\hspace{#2}}}

\newcommand{\sect}[1]{\setcounter{equation}{0}\section{#1}}

\begin{document}
\begin{titlepage}
\begin{flushright}
MIT-CTP-2708\\
\end{flushright}
\LARGE
\vspace{0.3in}

\begin{center}
{\bf Supersymmetric Partners of Oblique Corrections } \\
\vspace{0.4in}

\normalsize

{Emanuel Katz, Lisa Randall and Shufang Su}\\
\vspace{.05in}
{ \small \it Center for Theoretical Physics\\
Laboratory for Nuclear Science and Department of Physics\\
Massachusetts Institute of Technology\\
Cambridge, MA 02139, USA }
\end{center}
 
\vspace{0.2in}
 
\vspace{0.25in}
\LARGE
\begin{center}
Abstract
\end{center}
\normalsize

We discuss the violation of the equality of the gauge coupling  and
gaugino coupling in the presence of soft supersymmetry breaking.
Although this is a hard supersymmetry breaking effect, there are
finite contributions to this difference which can be determined as a
function of the soft supersymmetry breaking masses. The largest sources
of this difference are the ``super-oblique" corrections, which can be
enhanced by a multiplicity factor and a logarithm of the soft
supersymmetry breaking mass.  This is to be contrasted to standard
oblique corrections to the electroweak sector which decouple for large
supersymmetry breaking.   We show that these parameters
can be relatively large, particularly in models
of supersymmetry breaking motivated by solving the supersymmetric
flavor problem. We also perform a detailed study of the
non-oblique corrections for the example of squark decay.
We find that they can be significant, and should be accounted
for in the theoretical prediction.  

\end{titlepage}

\sect{Introduction}
\label{introduction}
Supersymmetry has yet to be discovered.  If/When it is, the
information about the underlying theory will be limited but powerful.
For example, the spectrum of superpartners might tell us about the
mechanism of communication of supersymmetry breaking. It  is
conceivable that there is very large supersymmetry breaking within
multiplets and that there are many as-yet undetected multiplets
carrying standard model gauge charge. Although the heavy
superparticles will elude direct detection, there can be large virtual
effects which could be detectable through the difference in the gauge
and gaugino coupling.

In this paper, we will discuss a new category of precision
measurements which one might use to probe supersymmetric theories. As
we will discuss in detail in Sec.~\ref{obli_and_superobi}, these
measurements have much in common with the well-studied ``oblique''
corrections which characterize corrections to the standard model.  In
supersymmetry, the gauge coupling $g$ is equal to the gaugino Yukawa
coupling $\tilde{g}$ when supersymmetry is exact.  However, when
supersymmetry is broken, components of the supermultiplet (for
example, \quark\ and \squark) no longer have the same masses, which
leads to  different contributions to gauge and gaugino propagators.  This
is the supersymmetric analog of the ``oblique'' corrections and gives rise 
to finite differences between $g$ and $\tilde{g}$.   
The experiments will certainly be more difficult because  they
require very precise measurements and good measurements of 
  other supersymmetric
parameters. However, these measurements could yield invaluable 
insights into the
underlying physics of supersymmetry breaking.

The key to understanding the quantities we wish to consider is to
distinguish hard and soft supersymmetry breaking parameters.
Supersymmetry can only be broken softly  if it is to serve as a
solution to the hierarchy problem.  The focus of most research on
supersymmetric phenomenology is on the soft supersymmetry
breaking parameters, and in particular the superpartner masses. Hard
supersymmetry breaking, on the other hand, is ``forbidden''.  However,
this is not true, even in the softly broken supersymmetric theory!
The only constraint is that the  coefficients of hard supersymmetry
breaking operators are {\it finite}.  There are no available
counterterms in the supersymmetric Lagrangian.  Nonetheless, finite
corrections are permissible and present. This is precisely analogous
to the precision electroweak parameters $S$, $T$, and $U$ which are
parameters of the effective theory which can be calculated as a
function of the underlying heavy masses. In fact, our focus here is on
new parameters which reflect hard supersymmetry breaking in the form
of oblique corrections in the gauge sector as well. As with the
precision electroweak parameters, these parameters will serve as a
test of the supersymmetric model, and give insights into non-standard
physics. While the standard oblique corrections were sensitive to
large violations of custodial SU(2) symmetry, the super-oblique
corrections are sensitive to large splittings within a supersymmetric
multiplet.

The largest effects will occur if some superpartners have
multi-TeV masses, though there are potentially measurable
effects even with smaller mass splittings. Examples of
models with large mass splittings are given in 
 \cite{ckn,dp,dg,mi} in which the partners of the first two
generations are heavy, suppressing  potential flavor changing neutral
current (FCNC) problems.   The squark masses could also be heavier
than the slepton masses because of the larger contribution from the
running of the masses due to the SU(3) strong coupling constant,
or because of a larger tree-level mass in a gauge-mediated model.
In Sec.~\ref{numbers} we present three different models which give
different predictions of the super-oblique corrections motivated by
these considerations and compare the effects.

Oblique corrections are also important because they are process
independent, and therefore can be constrained by different
experiments. Non-oblique corrections on the other hand
depend on the particular process under consideration. Furthermore,
virtual heavy states decouple and will not give large effects.
Finally, there is no multiplicity factor (if there are many
multiplets with large splittings). However, it is possible
that non-oblique supersymmetric loop corrections could
be important  when extracting the oblique
parameters. This is because of the enhancement by a group theory factor
(from diagrams with internal gauge bosons and gauginos).  Since
the dominance of the oblique corrections requires
that these effects are small, we have also calculated non-oblique
corrections. These include vertex, wave function renormalizations,
and gluon emission. A detailed check on our calculation
is the cancellation of infrared divergences, as is discussed
in Appendix~\ref{appE}.
 For simplicity, in this paper, we apply our calculation of non-oblique
corrections only to squark decay; we find the non-oblique corrections
can in fact be significant. This indicates that it will be important
to also incorporate non-oblique corrections in order to extract
the masses of heavy parameters dominating the oblique-corrections. 

Our work is in some ways complementary to the recent papers
by  Cheng, Feng, and Polonsky\cite{cfp2} and Pierce, Nojiri,
and Yamada\cite{npy}
who considered in greater detail methods for extracting oblique
parameters from experiments. We instead focus on the theoretical
``uncertainty" arising from non-oblique corrections which
could corrupt the extraction of parameters if they are large.
Of course, these can ultimately all be incorporated.
We also add to the consideration of experimental signatures
by giving the prediction for the production of $b \tilde{b} \tilde{g}$
and $t \tilde{t} \tilde{g}$ (assuming it is kinematically accessible)
which could prove to be a promising method of extracting SU(3) gaugino 
Yukawa coupling
if the supersymmetric partners are light.

We begin in Sec.~\ref{obli_and_superobi} discussing the analogy
between the super-oblique and oblique corrections and the
non-decoupling effect of heavy superparticles.  In Sec.~\ref{FCNC},
we present   different examples of models  that contain 
heavy superpartners.
We give the results for super-oblique
correction for these models in Sec.~\ref{numbers}.  
In Sec.~\ref{non_stan}, we  consider additional gauge representations
and their    
contribution to the super-oblique corrections.  In
Sec.~\ref{measure}, we  discuss possible
ways to measure the   super-oblique parameters and give
the results for the $b \tilde{b} \tilde{g}$ and $t \tilde{t} \tilde{g}$
production cross sections.
In Sec.~\ref{non_obli}, we investigate the non-oblique corrections
and show that they could be quite significant compared to the oblique 
corrections.    In Sec.~\ref{conclusion}, we conclude.
We have several appendices with details of our calculations.
Appendix A contains the $O(\alpha)$ corrections to vertices
and self-energies for arbitrary parameters. Appendix B gives
the super-oblique corrections.  Appendix C gives
the results of gluon emission, which will be necessary to cancel
infrared divergences. In Appendix D, we consider the results
in the limit of exact supersymmetry. Appendix E shows that
the results of Appendices A and C combine to give infrared finite results.
The final appendix gives the results for super-oblique corrections
from a messenger sector.
\nopagebreak
\sect{Oblique and Super-Oblique}
\label{obli_and_superobi}
It is useful first to review standard oblique corrections.  Precision
electroweak corrections were important in that they tested
consistency of the standard model and gave insight into the
high-energy world.  The oblique corrections are the corrections to
the gauge boson propagator. The parameters which characterize the
most important effects can be obtained by retaining the leading
pieces in a derivative expansion, accounting for gauge
invariance. The six ``parameters'' are $\Pi'_{\gamma \gamma}$, $\Pi'_
{\gamma Z}$, $\Pi_{WW}$, $\Pi'_{WW}$, $\Pi_{ZZ}$, $\Pi'_{ZZ}$.  We
chose to work in the basis of tree-level mass eigenstates and
retained only the nondecoupling pieces of the propagator, those of
dimension four or less.     It is useful  
to absorb three of the parameters into $g$, $g'$, $v$, which leaves
 three new finite parameters which are forbidden by gauge
invariance but  are permitted once it is spontaneously broken.   In
Ref.~\cite{gr}, these three parameters  were identified as as $\delta
Z_{ZZ}$, $\delta Z_{\gamma Z}$, $\delta m_{ZZ}$.  In the more widely
utilized Peskin-Takeuchi naming convention, they were absorbed into
parameters \cite{pt1,pt2}  $S$, $T$, $U$, while a third popular
convention from a paper of Altarelli and Barbieri \cite{ab} absorbed
them in the three parameters $\epsilon_1$, $\epsilon_2$,
$\epsilon_3$.  The importance of these parameters is that even heavy
states do not decouple.  For example,  one can compute the
contributions of a heavy doublet to the  standard $S$ parameter, 
 \beq \Delta S={1 \over 6\pi} \left[1-Y 
\log ({m_{\rm N}^2 \over m_{\rm E}^2}) \right] .   
\eeq

Of course, the many new particles present in a supersymmetric theory
can contribute to the standard oblique corrections.  In general, these effects
are   small because heavy superpartners decouple.
 However,  there is a potentially more interesting
class of ``oblique" corrections for supersymmetric theories.
Supersymmetry guarantees equality of the gauge and gaugino Yukawa
couplings.  When supersymmetry is broken, these are no longer
guaranteed to be equal at low energy.

We   consider the relative difference of the gauge and
gaugino Yukawa couplings, which is sensitive to supersymmetry
breaking.  Due to supersymmetry, this difference is UV-finite and
hence independent of the renormalization scale $\mu$.   
This difference  could be  measured by comparing the Yukawa coupling
$\tilde g$,  extracted from some physical process at energy scale
$q^2$, to the gauge coupling $g$ at that same energy.    Because $\alpha$ is
reasonably small, and logarithms are not large, we choose to  
work to leading order in $\alpha$, rather than taking an effective
theory approach and performing a full leading log calculation. 
This is useful since the finite pieces can be competitive
with the logarithmically enhanced terms, while higher
order terms are negligible.   Note also
that these parameters are $\mu$-independent and well defined, so one
can calculate  both logarithmically enhanced
and nonlogarithmically enhanced   contributions.

We define the gauge coupling $g(q^2)$, and the gaugino Yukawa coupling
$\tilde g(q^2)$, as the couplings containing all the ``super-oblique''
corrections; in other words, as the couplings containing contributions
to the gauge (or gaugino) propagator from all possible squarks and
quarks running in the loops.  We note that our definition of $g(q^2)$
(which is process independent) differs from the  convention of
Ref.~\cite{feng}, who  used an effective theory language and therefore
did not include  in their definition
oblique  contributions from light fermion loops.
Our two conventions differ by these finite
contributions, which are generally smaller than the logarithmic
contribution but not by a substantial amount. Since one can readily
calculate these additional loops, which the full order $\alpha$
calculation would require in any case, we choose to include these
pieces also in our definition of oblique corrections.

Hence, in our case the difference between the couplings will receive
contributions from all ``super-oblique'' diagrams (including the light
fermion loops).  These consist of the quark and squark contributions
to the the gauge propagator (Fig.~\ref{wgluon} (3) and (4)), and the
gaugino propagator (Fig.~\ref{wgluino} (2)).  For general choices of
superpartner masses, the difference at one loop takes the form of a
function
\beq \frac{g(q^2) - \tilde g(q^2)}{g(q^2)}  =
\frac{\alpha}{4\pi} F_{\rm ob}\left(\frac{m_{q_{\rm i}}^2}{q^2},
\frac{m_{\tilde{q_{\rm i}}}^2}{q^2},
\frac{m_{\tilde
g}^2}{q^2}\right)=\frac{\alpha}{4\pi}\frac{F_{g3}
+F_{g4}-F_{\tilde{g}2}}{2},
\label{defob}
\eeq 
which depends on the ratios of the various masses with the  
scale
$q^2$ ($F_{g3}$, $F_{g4}$, $F_{\tilde{g}2}$ are explicitly  given in
Appendix~\ref{appA} as integrals over Feynman parameters).  For
example, in the case of the SU(3) gauge and gaugino Yukawa couplings,
the contribution of a single heavy doublet superfield to this
difference at $q^2 \sim m_{\tilde g}^2$ is
\begin{equation}
{g_3-\tilde{g}_3 \over g_3}={\alpha_{3} \over 24 \pi}\left( {\rm Log}
\left({m_{\tilde{g}} \over m_{\tilde{q}}} \right)^2
-{11 \over 12}\right).
\label{simpli_ob}
\end{equation}
This formula is only true in the limit $\mq^{2}\ll{q}^{2}$ and
$q^{2}\ll\msq^{2}$.  A detailed derivation of this formula 
is given in Appendix~\ref{appB}.  Also notice that  because
physical quarks  are generally lighter than the gauginos, the
logarithm is cut off  at the gaugino mass. Therefore, the logarithm
is only large if there is a sizable splitting between the squark or
slepton and gaugino masses.

Although the oblique corrections  first arise at one loop (and there is
an  additional 1/6 suppression compared to the usual
$\alpha/(4\pi)$), there are factors which can make it large.  First of
all, we see that there is a logarithm; if there is a large mass
splitting, there will be  a corresponding enhancement.  Second, there
is a multiplicity factor; ``every" particle with gauge charge
contributes.  The logarithm here  differs from standard oblique
corrections in that  decoupling works differently.  Heavy scalar
superpartners contribute {\it more} rather than less.

If we call the soft scalar mass $m_0$ and the coefficient of the
trilinear scalar coupling $A$, for the standard oblique corrections
(e.g. the $S$ parameter), it is easy to see that
\beq
\Delta S \propto{m_{\rm LR}^2 \over m^2}\sim {(\lambda_q A m_q )^2\over  
m_0^2},
\eeq
which decouples for large $m_0$.

In the case of the super-oblique corrections which we discuss, one
finds \beq \delta g(m_q) \propto {\rm Log}{m_{\tilde{q}}^2 \over  m_q^2},
\eeq (where we now assume the quark mass is bigger than the gaugino's
and take $q^{2}\sim\mq^{2}$) which clearly does not decouple, and is
in fact logarithmically enhanced.  This logarithm can however be
readily understood \cite{nojiri}.  If we assume a chiral multiplet in
which the scalar partner is heavy, there is an energy regime between
the fermion (or gaugino) mass and the heavy scalar mass in which the
gauge and gaugino coupling run differently.  In fact one can also
determine that when the scalar is the heavy partner, the gauge
coupling at low energy will be reduced relative to the gaugino
couplings due to the sign of the fermion contribution to the $\beta$
function for the gauge coupling.

Decoupling does occur  however for super-oblique corrections when
$m_q\sim m_{\tilde{q}}$ \beq \delta g \propto {(m_{\tilde{q}}^2-m_q^2)
\over m_q^2}.  \eeq Notice here that we have assumed the squark and
quark to be nearly degenerate and heavier than the gaugino.  Therefore
what is relevant to super-oblique corrections are standard model gauge
charge multiplets with large supersymmetry breaking splitting in  the
masses.  If these mass splittings are not  very large, the constant
term $11/12$ in Eq.~\ref{simpli_ob}  can be somewhat significant.
For instance,  if  $m_{\tilde q}=5m_{\tilde g}$ the constant term is a
$30 \%$ correction to the logarithm.

We conclude that heavy scalar partners can lead to interesting deviations
from the exact supersymmetric prediction $g=\tilde{g}$.

\sect{Large Effects and FCNC}
\label{FCNC}

It is reasonable to ask whether there is any reason to believe that
there are large mass splittings.   The answer might be yes, 
given the current ideas for resolving
the FCNC problems in supersymmetric theories.  The apparent degeneracy
of squarks or alignment with quarks is still not understood.  There are
two types of scenarios which offer the possibility of large mass  
splittings
among the superpartners. It could
be that squarks (and gluinos) are heavier because
of the larger SU(3) coupling. Alternatively,
a suggested resolution of the FCNC problem involves
heavy first two generations   \cite{ckn,ckn1,dp,dg}.

An example of a class of models in which the colored
states are heavy is the gauge-mediated models
which have been suggested as a resolution of the FCNC problem.
It could also be that colored scalars get enhanced
masses from renormalization group running. Of course,
either of these scenarios would probably make both the squarks
and gluino heavy. A very heavy gluino would not be subject
to the experimental tests  we are discussing, although
there could still be measurable oblique corrections for the SU(2)
and U(1) sectors.  Just to test how big the effects can reasonably
be expected to be in a model in which the differences between 
the gauge coupling and gaugino Yukawa coupling can all
be measured, we took a toy model with heavy squarks and a relatively
light gluino. Although not motivated by any particular model,
it serves to illustrate the potential size of the oblique corrections.

There are several groups who investigated the possibility
of heavy first two generations. An example of such a model is \cite{dp},
where an anomalous U(1) symmetry which served as a mediator of supersymmetry
breaking was introduced. If only the first two generations are charged,
their mass can be of order $4 \pi$ bigger than the lighter
third generation states. The gauginos are also among the light states.
 In Ref.~\cite{dg}, the heaviness of the first two generations
is achieved through the constraints of naturalness with some
restrictions on  fine tuning. All the superparticles can be divided
into two groups.  The first, called ``the brothers of the Higgs'',
contains superpartners whose
masses will strongly influence the mass running of the Higgs and  thus
are strong constrained by naturalness.  The second, called
``the cousins of the Higgs'', consists of superparticels
which are less constrained and
hence can be heavy.  The first two generation squarks are in the latter
group and their mass can be in the range of 900 GeV$-$5 TeV.
Related ideas are incorporated into the  ``More Minimal Supersymmetric
Standard Model'' (also called effective supersymmetry)\cite{ckn}, in which one
attempts to allow the maximal masses consistent with naturalness
bounds.  In this model, a new gauge group $G$ is introduced, which
enlarges the accidental symmetry group and thus forbids the
renormalizable $B$ and $L$ violating interactions.  It also introduces a
new mass scale $\tilde{M}\sim5-20$ TeV which sets the mass scale of
the first two generations.  This model permits the first two
generations of squarks to be heavier which goes a considerable way
towards solving flavor changing problems.  In addition, the
requirement of naturalness implies that some squarks ($\tilde t_{\rm L},
 \tilde{b}_{\rm L}$) and most gauginos must have a mass 
below $\sim 1$  TeV.  One
can then imagine $\tilde{m}_{1,2}\sim 20$ TeV, $\tilde{m}_3 \sim
100$ GeV $-$ 1 TeV.

\sect{Parameters and Numbers}
\label{numbers}

One can now ask for the
magnitude of the finite difference between the gauge and Yukawa
couplings for the kinds of models described above.  There are
three measurable quantities associated with $g_3$, $g_2$, $g_1$, which
are the coupling constants of the gauge groups SU(3), SU(2) and
U(1) respectively. These fit in well with a modified Peskin-Takeuchi
naming convention $v=(g_3-\tilde{g}_3)/g_3$,
$w=(g_2-\tilde{g}_2)/g_2$, $y=(g_1-\tilde{g}_1)/g_1$.  

We now present the magnitude of each of these quantities for the
types of models described in Sec.~\ref{FCNC} in which there are mass
splittings.  We make simple assumptions below to get an idea of the
magnitude of the effects.  In the following we will use the value of
the various couplings at $M_Z$ to estimate the effects, and we take
$q^{2}\sim{m}_{\rm gaugino}^{2}$. 

We define three  different models:
\begin{description}
\item[Model 1]   Only the squarks are heavy. 
We take
$R_{\tilde{q}}^{\tilde{W}}\equiv{m}_{\tilde{W}}/m_{\tilde{q}}=0.05$, and
since this is a gauge mediated model
$m_{\tilde{g}}:m_{\tilde{W}}=\alpha_3:\alpha_2\approx7:2$,where
$\alpha_3=0.12$, $\alpha_e=1/128$, $\sin^2\theta_W=0.23$.  Thus,
$R_{\tilde{q}}^{\tilde{g}}\equiv{m}_{\tilde{g}}/m_{\tilde{q}}=0.18$.  
In Table~\ref{model1}, we present the numerical results for Model 1. 
\begin{table}
\begin{tabular}{ccc}\hline
$\hspace*{0.9in}v\hspace*{0.5in}$&$3.82\%\log{R}_{\tilde{q}}^
{\tilde{g}}-1.8\%$&\hspace*{0.5in}-8.4\%\hspace*{0.9in}\\
$\hspace*{0.9in}w\hspace*{0.5in}$&$0.81\%\log{R}_{\tilde{q}}^
{\tilde{W}}-0.37\%$&\hspace*{0.5in}-2.8\%\hspace*{0.9in}\\
$\hspace*{0.9in}y\hspace*{0.5in}$&$0.30\%\log{R}_{\tilde{q}}^
{\tilde{W}}-0.14\%$&\hspace*{0.5in}-1.0\%\hspace*{0.9in}\\ \hline
\end{tabular}
\caption{Super-oblique corrections for Model 1.}
\label{model1}
\end{table}
Here we have ignored the additional finite contribution from the sleptons.
If, for example, $m_{\tilde{l}}=m_{\tilde{W}}$, one calculates
the finite contribution $\alpha_2/24\pi(11/2-\sqrt{3}\pi/2)$ instead of 
$\alpha_2/24\pi(-11/12)$ in 
Eq.~\ref{simpli_ob} 
(where we consider SU(2) case, and similarly for U(1) case).  

\item[Model 2]   A More Minimal Spectrum, 
in which the masses of $b_{\rm R}$, $u$,
$d$, $s$, $c$, $e$, $\mu$ are $m_{\rm H}=20$ TeV, whereas the masses of
$b_{\rm L}$, $t$, $\tau$ are $m_{\rm L}=$1 TeV, while $m_{\tilde{W}}
=100 {\rm GeV}$,
and $m_{\tilde{g}}=350{\rm GeV}$.  In this model,
$R_{\rm L}^{\tilde{g}}\equiv\mgluino/{m}_{\rm L}=0.35$,
$R_{\rm H}^{\tilde{g}}\equiv\mgluino/{m}_{\rm H}=0.018$,
$R_{\rm L}^{\tilde{W}}\equiv{m}_{\tilde{W}}/{m}_{\rm L}=0.1$ and
$R_{\rm H}^{\tilde{W}}\equiv{m}_{\tilde{W}}/{m}_{\rm H}=0.005$.   
In Table~\ref{model2}, we present the numerical results for Model 2. 
\begin{table}
\begin{tabular}{ccc}\hline
$\hspace*{0.6in}v\hspace*{0.2in}$&$0.95\%\log{R}_{\rm L}^
{\tilde{g}}+2.86\%
\log{R}_{\rm H}^{\tilde{g}}-1.8\%$&\hspace*{0.2in}-14.3\%
 \hspace*{0.5in}\\
$\hspace*{0.6in}w\hspace*{0.2in}$&$0.36\%\log{R}_{\rm L}
^{\tilde{W}}+0.72\%
\log{R}_{\rm H}^{\tilde{W}}-0.5\%$&\hspace*{0.2in}-5.1\%
 \hspace*{0.5in}\\
$\hspace*{0.6in}y\hspace*{0.2in}$&$0.16\%\log{R}_{\rm L}
^{\tilde{W}}+0.3
8\%\log{R}_{\rm H}^{\tilde{W}}-0.25\%$&\hspace*{0.2in}
-2.6\% \hspace*{0.5in}\\ \hline
\end{tabular}
\caption{Super-oblique corrections for Model 2.}
\label{model2}
\end{table}
\item[Model 3] A modified more minimal model similar to Model
2, in which the sleptons have mass $m_{\rm L}=$1 TeV.
In Table~\ref{model3}, we present the numerical results for Model 3. 
\begin{table}
\begin{tabular}[t]{ccc}\hline
$\hspace*{0.6in}v\hspace*{0.2in}$&$0.95\%\log{R}_{\rm L}
^{\tilde{g}}+2.86\%
\log{R}_{\rm H}^{\tilde{g}}-1.8\%$&\hspace*{0.2in}-14.3
\%\hspace*{0.5in}\\
$\hspace*{0.6in}w\hspace*{0.2in}$&$0.54\%\log{R}_{\rm L}
^{\tilde{W}}
+0.54\%\log{R}_{\rm H}^{\tilde{W}}-0.5\%$&\hspace*{0.2in}
-4.6\%\hspace*{0.5in}\\
$\hspace*{0.6in}y\hspace*{0.2in}$&$0.32\%\log{R}_{\rm L}
^{\tilde{W}}
+0.22\%\log{R}_{\rm H}^{\tilde{W}}-0.25\%$&\hspace*{0.2in}
-2.2\%\hspace*{0.5in}\\ \hline
\end{tabular}
\caption{Super-oblique corrections for Model 3.}
\label{model3}
\end{table}
\end{description}
We can see that $v$ is the same for Models 2 and 3 since changing the 
slepton mass would not affect the SU(3)
coupling constant, which only gets contributions from \quark, 
\squark\
and \gluino.  $v$ is much larger than $w$ and $y$ since
the corresponding coupling  is much larger.
We write the super-oblique corrections in terms of the ratio of the
mass splittings in Tables~\ref{model1}, \ref{model2}, \ref{model3}, 
where it is evident that
the finite piece 11/12 does give a sizable contribution if the logarithm
is not much larger than 1.
The results we get for Model 1 are consistent with the results for 
the ``heavy QCD
models'' in Ref.~\cite{feng2} when the gluino is light.  Our model 2 is
close to the ``2-1 models'' \cite{feng2} except that we put $b_{\rm R}$ in the
heavy sector.  When we let $b_{\rm R}$ also be light, the coefficient
before $\log{R}^{\tilde{W}}_{H}$ in Model 2 also 
fits well with the results given in
Ref.~\cite{feng2}.  However, in our case, we also calculate the
contribution from the finite piece 11/12 and the light squarks,  
whose masses
are still large enough (compared to the gauginos) to give noticeable contribution.

\sect{More Non-Standard Physics}
\label{non_stan}

In this section, we consider the role of the super-oblique 
correction measurements not only as a test of  the parameters of
the minimal extension of the standard model
but also as a potential  probe of other new physics.  We have
already stressed the fact that heavy squarks  can be large.
This is especially important
as it is precisely when the squarks become kinematically
inaccessible that their virtual effects become largest.
If one does not find the squarks, or if they are at the
edge of the kinematically accessible regime, one would
like to be able to confirm their existence and attempt
some estimate of their mass through their virtual effects.

However, it might be that the squarks are relatively light, and one
still measures large deviations from the supersymmetric
relations. Provided they are not too large, one might ask how this
could be consistent with supersymmetry.  In fact, this is probably the
most exciting possibility.  It would indicate the existence of
nonstandard fields which participate in supersymmetry breaking and
therefore have a large mass splitting relative to the mass of the
fermions or scalars and which carry standard model gauge charge. An
obvious candidate in terms of existing models is a messenger sector,
although possibilities extend beyond this.  As an example, suppose the
messenger sector consisted of a  large vectorlike representation under
SU(5). One should note that for the ``standard'' messenger scenario,
with a singlet coupled to messengers through a Yukawa term,  the
masses of the scalars are $\sqrt{\lambda^2 S^2\pm \lambda F_{\rm S}}$
whereas the fermion mass is $\lambda S$.   In this case, $q^{2}$ (the
external four-momentum squared for $g$ and \gluino) is taken to be smaller
than the fermion mass.  We take the limit
$q^2\ll{m}_{\rm fermion}^2$, $q^2\ll{m}_{\rm scalar}^2$ to calculate the
super-oblique corrections.  Let $m_{\rm scalar}^2/m_{\rm fermion}^2=1\pm{x}$.
The effects are unfortunately small unless $x$ is very close to 1
(see Appendix~\ref{appF}).
 Without an enormous multiplicity factor,  
a   nontrivial mass splitting is needed  to get a
sizable contribution.

\sect{Measurements of Oblique Parameters}
\label{measure}

It is clear that there can be fairly sizable effects. It is therefore
extremely interesting to ask the question whether the parameters
can be measured. The numbers presented above serve as  
benchmarks for
interesting measurements, although a measurement at  any level of
accuracy will constrain our confidence in supersymmetry or
equivalently constrain extensions of the minimal scenario.

Several groups have begun to investigate this question.  Initially
this was pursued as a test of supersymmetry at tree level \cite{feng}
and it was later  studied in order to extract squark masses if
they are heavy \cite{nojiri,cfp2}. More recently,
several groups, \cite{us,feng,feng2,npy} have
studied super-oblique corrections. We summarize some suggestions for measurements. We also
  present new results for the $b\tilde{b} \tilde{g}$ and 
$t \tilde{t} \tilde{g}$ productions.

One measurement which has been suggested of the U(1)
coupling is in a paper by Nojiri, Fujii, and Tsukamoto \cite{nojiri}
where they studied $\sigma(e^+ e^- \to \tilde{e}_{\rm R}^+
 \tilde{e}_{\rm R}^-)$.
This process proceeds through $s$-channel gauge exchange and
$t$-channel neutralino exchange. One can measure $\tilde{g}_1$ and
$M_1$ by measuring the angular distribution of the differential cross
section ${d\sigma(e^+ e^- \to \tilde{e}_{\rm R}^+ \tilde{e}_{\rm R}^-)}/{d
\cos\theta}$.  It has also been suggested
by Cheng, Feng, and Polonsky \cite{cfp2}
to look at the $e^- e^-$ scattering cross
section through  $t$-channel neutralino exchange
when running the collider in the appropriate mode.  This can
lead to even greater precision since some of the troublesome backgrounds
are absent and the ability to highly polarize the beams increases the
$\tilde{e}_R\tilde{e}_R$ cross section further \cite{feng,feng2}.
It should be borne
in mind that although the U(1) coupling  might be the  most accurately
measured, the magnitude of $y$ is the smallest of the three precision
measurements.  It is therefore worthwhile to investigate the
possibility of measuring $w$ and $v$ as well.

Several ways of measuring  $w$ have been suggested; the most
appropriate will depend on the region of supersymmetry parameter space
which is realized  
\cite{feng,cfp2}. We summarize some of the suggestions
of Ref.~\cite{cfp2}.
If the charginos are kinematically accessible, their production
cross section can be used to measure $w$, for example if they are pair
produced with a contribution from sneutrino exchange. 
Neutralino production (and decay
if more than one mode is sizable) can also be used to test $w$ (and $y$).
It might also be possible to test  the gaugino coupling   
through the chargino mass dependence on the $H \tilde{H}
\tilde{W}$ vertex, though this will  help only away from the region
where charginos are nearly pure gaugino.
It should be noted that a  good measurement of $m_{\tilde{\nu}}$  
will be important
when constraining  $w$  to high precision by some of these methods.  
Another possible
way to study $w$ is through the branching fraction of charginos
if there are appropriate modes which are kinematically
accessible. Finally, selectron and sneutrino production could
provide important probes.

Finally we consider the measurement of $v$, which requires measuring
the SU(3) Yukawa coupling.  It is  very likely that  the best test
will be at a hadron collider, where gluino and squark can be produced
abundantly.  In the case that squark decay is observed, the branching
ratios into gluinos and winos can serve as probes of the super-oblique
corrections. Notice that the gluino production section
goes primarily through the gluon coupling, and therefore
does not serve as a test at all. The branching fraction
tests are also possible at an $e^+e^-$ collider if squarks
are sufficiently light.

 At an $e^+ e^-$ collider, it might be possible to test
the 2-1 type models if  $\tilde{t}$, $\tilde{g}$ are sufficiently 
light\footnote{We thank Michael Peskin for suggesting this possibility} or
if $\tilde{b}$, $\tilde{g}$ are sufficiently light\footnote{We thank
Jonathan Feng for pointing out this alternative possibility.}.  We  consider
the  the cross section for
$t \tilde{t} \tilde{g}$
and $b \tilde{b} \tilde{g}$ production to get an estimate of the
possible sensitivity to $v$.  For $b$ quark and squark
production one predicts a few hundred events according to the gluino
and squark mass (below  a few hundred GeV) and roughly the same for
the top quark.  In Fig.~\ref{eet} and Fig.~\ref{eeb} we present the numbers
of events produced per year at an  $e^{+}e^{-}$ collider at $E_{\rm cm}=1000$
GeV and luminosity $L=5\times10^{33} {\rm cm}^{-2}{\rm s}^{-1}$.  
The number of  
events
decreases with the increase of $\msq$ and $\mgluino$ due to the
decrease in phase space.  The number of $b$ events  is slightly less
than $t$ events, though their phase space is larger.  This is because
of the different couplings of the $b$ and $t$ to $Z$ and $\gamma$.
When $\msq>\mgluino$, it is possible that the intermediate squark state
is on shell, which gives a peak in the   production rate
which we have not shown.  This   is 
more  significant to $b\tilde{b}\tilde{g}$ production since $m_{b}\sim{4}$ GeV,
 so there is a larger range of parameters
for  which the intermediate $\tilde{b}$  is on shell. 
In this case one is in fact measuring the squark decay branching
ratio into gluino, which can also determine the gaugino Yukawa
coupling.   By measuring the  branching ratio for the decay
of the   \squark\   into
$\tilde{q}\rightarrow{q}\tilde{g}$,  
$\tilde{q}\rightarrow{q'}\tilde{\chi}^\pm$
and $\tilde{q}\rightarrow{q}\tilde{\chi}^0$, one
can hope to extract the coupling, if the other parameters
(masses) can be independently determined \cite{feng2}. The number of events we present
here is the event production, not corrected for cuts
necessary to remove background events. Nonetheless for sufficiently
light gaugino and third generation squarks, it seems likely that a
measurement of the gaugino coupling at at least the 15\% level should
be possible.

\sect{Non-oblique Contributions}
\label{non_obli}
When extracting the \squark\quark\gluino\ coupling from a physical
process, such as squark decay  branching fractions,
it is important to consider finite
contributions from non-oblique diagrams.  Although one usually ignores
such terms, they could be significant because of the $C_2(G)$ factor
($C_{2}(G)=N$ for SU(N)).  For instance, in the non-abelian gluon
and gluino loop diagrams (Fig.~\ref{susyv} (4)), even though there
are no large logarithms, the $C_2(G)$ acts as a multiplicity factor,
enhancing their effect. Hence, it is useful to check the
relative size of the non-oblique corrections.

 It is important to bear in mind that the $\mu$ dependence
in the oblique parameters cancel; everything is finite. We will sometimes
refer to finite pieces; by this we mean the nonlogarithmically
enhanced contribution (that is, the finite pieces in an effective
theory approach).  The finiteness is important in that
it means that the answer at one-loop is renormalization scheme independent.
The dependence on momentum is through the dependence on the physical
momenta which enter the loops. For any given process these are specified.
For example, in the case of squark decay, the mass squared of the
squark will enter.  In a process with a virtual squark, the loops
will be functions of $q^2$ and the squark mass squared; the $q^2$
dependence  of the virtual corrections can be incorporated
when one integrates over $q^2$.  Because it is clearly simpler,
we illustrate the non-oblique corrections (which are process dependent)
with the specific case of squark decay.  One can however apply
our results to any process by which the oblique parameters will be measured. 

Now let us focus on the \squark\ decay process
$\tilde{q}\rightarrow{q}\tilde{g}$.  The relevant non-oblique diagrams
include the quark and squark self energies (Fig.~\ref{wquark},
Fig.~\ref{wsquark}), the non-abelian gluon and gluino self energies
(Fig.~\ref{wgluon} (1), (2) and Fig.~\ref{wgluino} (1)), the
$\tilde{q}q\tilde{g}$ vertex diagrams (Fig.~\ref{susyv}), and the
gluon emission diagrams (Fig.~\ref{emission}).  Here,  all the
diagrams are necessary to render a gauge invariant answer free of
infrared divergencies.  The quark and gluino are assumed to be always
on shell, while the squark could be offshell at energy $q^2$.  We note
that for completeness we have included both abelian diagrams
(Fig.~\ref{susyv} (1), (2), Fig.~\ref{wquark}, Fig.~\ref{wsquark})
(proportional to $C_2(N)$), and non-abelian diagrams
(Fig.~\ref{susyv} (3), (4), Fig.~\ref{wgluino} (1), Fig.~\ref{wgluon} (1), (2))
(proportional to $C_2(G)$). We use the dimensional reduction
regularization scheme ($\overline{DR}$), rather than dimensional
regularization, to avoid obtaining a finite difference between the
gauge and the Yukawa couplings in the supersymmetric limit.  This
finite difference occurs in dimensional regularization because of the
mismatch of fermionic and bosonic degrees of freedom for dimension $d>4$.

In addition, we introduce a gluon mass, $\lambda$, to regulate
infrared divergencies. Since the quark mass is usualy much smaller
than that of the squark and gluino, we could have set $m_q$ to zero.
However, we found that because $m_q$ effectively regulates the
collinear divergences in the angular integral of the real gluon
emission, keeping it non-zero made it easier to keep track of the
various sources of divergences.  As a result, separation of  soft
emission divergences, proportional to the logarithm of $\lambda$,
from those of collinearity (proportional to the logarithm of $m_q$)
was straightforward.  Identifying the infrared  divergence in a given
vertex diagram with the corresponding divergence in the emission
diagram, we were able to explicitly verify that the sum of diagrams is
infrared finite (see Appendix~\ref{appE}). The total UV-finite virtual
non-oblique correction is defined to be

\begin{equation}
F_{\rm virtual}(q^2, m_q^2, m_{\tilde q}^2, m_{\tilde g}^2, \lambda^2)
=\sum_{i=1}^{4}\tilde{F}_{\rm Vi}+\frac{1}{2}\sum_{{\rm i}=1}^{2}
F_{q{\rm i}}+\frac{1}{2}\sum_{
{\rm i}=1}^{2}F_{\tilde{q}{\rm i}}+\frac{1}{2}F_{\tilde{g}1},
\end{equation}
Here, the $\tilde{F}_{\rm Vi}$ are the vertex diagrams contributions,  and
$F_{q{\rm i}}$, $F_{\tilde{q}{\rm i}}$, $F_{\tilde{g}1}$ are the quark, squark,
and gluino self-energies. All the $F$s are  explicitly given in terms
of integrals over Feynman parameters in Appendix~\ref{appA}.  We also
define the UV-finite contribution arising from gluon emission,
$F_{\rm emission}$,  by
\begin{equation}
\Gamma_{\tilde q \rightarrow q \tilde g g}(q^2) =  
\frac{\tilde\alpha}{4\pi}
\tilde\Gamma_0(q^2)F_{\rm emission}(q^2, m_q^2, m_{\tilde q}^2, 
m_{\tilde g}^2, 
\lambda^2).
\end{equation}
Here,  $\tilde\Gamma_{0}$ is the tree level decay rate for
$\tilde{q}\rightarrow{q}\tilde{g}$ and its explicit formula is given in 
Eq.~\ref{gamma0}.
Note that in the three-body final state decay we are integrating  
over all of
phase space, effectively including soft and hard gluon emission.  This
constitutes an overestimate of the emission finite contribution. However,
the sign is the opposite to the overall sign of the non-oblique
corrections, so this constitutes an underestimate of their size.

The total squark decay width is thus given by 
\begin{equation}
\Gamma_{\rm total}=\tilde\Gamma_{0}\left(1+\frac{\tilde\alpha}{4\pi}
(2F_{\rm virtual}+F_{\rm emission})\right).
\label{emi}
\end{equation}
Recall that in Eq.~\ref{defob}
\begin{equation}
\frac{g(q^2) - \tilde g(q^2)}{g(q^2)}=\frac{\alpha}{4\pi} F_{\rm ob},
\end{equation}
then 
\begin{equation}
\tilde{\alpha}=\alpha(1-\frac{\alpha}{4\pi}2 F_{\rm ob}).
\label{fob}
\end{equation}

Combining Eq.~\ref{emi} and \ref{fob} (We use $\alpha$ in stead of $\tilde\alpha$ in Eq.~\ref{emi} because the difference is high order.), the squark decay width is given in terms
of $g(q^2)$ as
\begin{eqnarray}
\Gamma_{\rm total}&=&\Gamma_{0}\left(1+\frac{\alpha}{4\pi}
(2F_{\rm virtual}+F_{\rm emission
}-2F_{\rm ob})\right)\nonumber\\
&=&\Gamma_{0}\left(1+\frac{\alpha}{4\pi}(2F_{\rm non-ob}
-2F_{\rm ob})\right),
\end{eqnarray}
where $F_{\rm non-ob}$ is defined by
$F_{\rm non-ob}=F_{\rm virtual}+1/2F_{\rm emission}$ and 
$\Gamma_{0}$ is similar to 
$\tilde\Gamma_{0}$ except that the coupling constant is $g$ instead of 
$\tilde{g}$.

The numerical results for
SU(3) $(\alpha/4\pi)F_{\rm non-ob}$ for on shell \squark\ decay is shown in
Fig.~\ref{changegluino}.  For a range of  $\mgluino/\msq$
 of $0.5-0.8$, which is a reasonable range for
the parameters which might be accessible to an $e^+ e^-$ collider, 
$(\alpha/4\pi)F_{\rm non-ob}$ changes from -8.5\% to -1.57\%.  
This can be a significant
fraction of $v$. Furthermore, it has the  
same sign to $F_{\rm ob}$ contributions,  thereby lessening the 
observed effect (Notice the relative minus sign between 
$F_{\rm ob}$ and $F_{\rm non-ob}$).
This cancellation is more significant when
$\mgluino/\msq$ is smaller.  Recall that we are
imagining a model like Model 2. For
 $m_{\tilde{g}}/m_{\rm L}=0.5-0.8$, with $m_{\rm L}/m_{\rm H}=1/20$, 
we obtain $v=-10.2\%\rightarrow
-9.7\%$ (where the parameter is given at $q^2=m_{\rm L}^2$).  

It might seem  surprising that the non-oblique corrections are larger 
in magnitude for smaller $\mgluino/\msq$, 
which seems  to contradict the decoupling effect of the heavy sectors.  
Actually this is due to some large negative total shift of 
the non-oblique corrections with respect to zero, 
while $F_{\rm non-ob}$ still goes
proportional to $\mgluino/\msq$, especially when the ratio is small, 
which shows the decoupling effect of the heavy sector.
  
$F_{\rm non-ob}$ for
SU(2) can  be obtained similarly, and is -1.43\% for
$m_{\tilde{W}}/m_{\tilde{l}}=0.5$ and -0.24\% for
$m_{\tilde{W}}/m_{\tilde{l}}=0.8$ (as compared to
$w=-2.6\% \rightarrow -2.4\%$).  A similar calculation for only
the abelian part is done in Ref.~\cite{Japanese}, which is consistent with our
result except that in their Eq.~(13), the constant in $F_{\rm real}$ should
be 7/2 instead of 13/4 and the $F_{\tilde{g}}$ in their Eq.~(19) should
have the opposite sign.

We only calculated the non-oblique corrections for on-shell squark decay.
For other physical processes, the non-oblique corrections 
can be done similarly using the same method. 
For example, in the process $e^+e^-\rightarrow\tilde{W}^+\tilde{W}^-$  
through t-channel $\tilde{\nu}$ exchange, one would also want to include
non-oblique corrections. However, there are additional radiation
diagrams which in general must be incorporated.   

As an additional check of our results, we have verified that in the SUSY
limit ($m_q = m_{\tilde q}, m_{\tilde g} =0$), the
non-oblique UV-finite corrections to $g_{\rm i}$ and $\tilde 
g_{\rm i}$ are identical
(See Appendix~\ref{appD}).

\sect{Conclusions}
\label{conclusion}
To conclude, it is obvious that if supersymmetry is discovered,
we would like to gain as many handles
on the underlying supersymmetric theory as can be
accessible. Super-oblique corrections provide a different perspective
into the physics of supersymmetry breaking and into the consistency of
the standard supersymmetric sector.  As with the standard
oblique corrections, they test consistency of the simplest
version of the theory and provide access to nonstandard physics.
They are most  useful
precisely when they are largest; namely there are heavy superpartners
which cannot be directly observed. The particular
parameters we discussed are the    supersymmetric analogs
of the standard-oblique parameters.  The one-loop gluon and gluino
propagators give contributions to the gauge and gaugino Yukawa
coupling difference, which are finite.   In nonstandard models,
especially those addressing the supersymmetric flavor problem, it is
likely that the parameters we have described could be large. 
We have shown that non-oblique corrections are also potentially significant,
and should be incorporated in the theoretical predictions.

The super-oblique parameters will be difficult
  to measure,
 although there are many potential candidate measurements.
Clearly, the better measured are the super-oblique parameters,
the more constrained will be the underlying model.   These parameters
merit further study, both in understanding experiments and
in further calculations of non-oblique effects.

\section*{Acknowledgments} We thank
Jonathan Feng, Ian Hinchliffe, Ken Johnson, and Michael
Peskin for useful  conversations.
 This research is supported in part by DOE under
cooperative agreement \#DE-FC02-94ER40818, NSF Young
Investigator Award, Alfred P. Sloan Foundation Fellowship,
DOE Outstanding Junior Investigator Award.

\appendix
\newcommand{\feynint}{{\int{{\rm d}x{\rm d}y{\rm d}z\,\delta(1-x-y-z)}}}
\newcommand{\wfeynint}{{\int{{\rm d}x{\rm d}y\,\delta(1-x-y)}}}
\newcommand{\kint}{\int_{\lambda}\frac{{\rm d}^{3}k}{(2\pi)^{3}2k_{0}}}

\sect{$O({\alpha})$ corrections to Vertices and Self Energies}
\label{appA}
In Fig.~\ref{smv} and \ref{susyv}, we show the one-loop Feynman
diagrams that contribute to the corresponding {\it gqq } and {\it
\~{q}q\~{g}} vertex corrections (denoted by $F_{\rm Vi}$ and
${\tilde{F}_{\rm Vi}}$).  Using the $\overline{DR}$ renormalization scheme
with $D=4-2\epsilon$ and regularizing the infrared divergences by an
infinitesimal gluon mass $\lambda$, we get\footnote{In all the
formulas and figures, we use SU(3) as an example.  For SU(2) and
U(1), the calculation is similar.}
\begin{eqnarray}
\label{F_{V1}}
\lefteqn{\hspace*{-0in}F_{\rm V1}=(C_{2}(N)-1/2C_{2}(G))
\left[\frac{1}{\bar{\epsilon}}+\log\mu^{2}-1-\right.}\nonumber\\
&&\hspace*{-0.2in}\feynint(2\log((x^{2}+2xy+y^{2})
\mq^{2}-xyq^{2}+z\lambda^{2})+\nonumber\\
&&\hspace*{-0.2in}\left.\frac{2(2-2x-2y-x^{2}-2xy-y^{2})\mq^{2}+
2(-1+x+y-xy)q^{2}}{(x^{2}+2xy+y
^{2})\mq^{2}-xyq^{2}+z\lambda^{2}})\right],\\
\label{F_{V2}}
\lefteqn{\hspace*{-0in}F_{\rm V2}=(C_{2}(N)-1/2C_{2}(G))\left[
\frac{1}{\bar{\epsilon}}+\log\mu^{2
}-\feynint\right.}\nonumber\\
&&\hspace*{-0.2in}(2\log((x+y)\msq^{2}+z\mgluino^{2}+
(-x-y+x^{2}+2xy+y^{2})\mq^{2}-xyq^
{2})-\nonumber\\
&&\hspace*{-0.2in}\left.\frac{4(x+y)(1-x-y)\mq^{2}}{(x+y)
\msq^{2}+z\mgluino^{2}
+(-x-y+x^{2}+2xy
+y^{2})\mq^{2}-xyq^{2}})\right],\\
\label{F_{V3}}
\lefteqn{\hspace*{-0in}F_{\rm V3}=C_{2}(G)\left[\frac{3}
{2\bar{\epsilon}}
+\frac{3}{2}\log\mu^{2}-
\frac{1}{2}-\feynint\right.}\nonumber\\
&&\hspace*{-0.2in}(3\log((-x-y+z+x^{2}+2xy+y^{2})\mq^{2}-xyq^{2}
+(x+y)\lambda^{2})+\nonumber\\
&&\hspace*{-0.2in}\left.\frac{3(1-2x-2y+x^{2}+2xy+y^{2})\mq^{2}
+(1+x+y-xy)q^{2}}
{(-x-y+z+x^{2}+
2xy+y^{2})\mq^{2}-xyq^{2}+(x+y)\lambda^{2}})\right],\\
\label{F_{V4}}
\lefteqn{\hspace*{-0in}F_{\rm V4}=C_{2}(G)\left[\frac{1}
{2\bar{\epsilon}}+
\frac{1}{2}\log\mu^{2}-
\frac{1}{2}-\feynint\right.}\nonumber\\
&&\hspace*{-0.2in}(\log(z\msq^{2}+(x+y)\mgluino^{2}-(x+y-x^{2}
-2xy-y^{2})\mq^{2}-xyq^{2
})-\nonumber\\
&&\hspace*{-0.2in}\left.\frac{\mgluino^{2}+(1-2x-2y+x^{2}
+2xy+y^{2})\mq^{2}
+xyq^{2}}{z\msq^{2}+
(x+y)\mgluino^{2}-(x+y-x^{2}-2xy-y^{2})\mq^{2}-xyq^{2}})\right],\\
\label{susyF_{V1}}
\lefteqn{\hspace*{-0in}\tilde{F}_{\rm V1}=(C_{2}(N)
-1/2C_{2}(G))\left[\frac{1}
{\bar{\epsilon}}+\log\mu^{2}-\frac{1}{2}-\feynint\right.}\nonumber\\
&&\hspace*{-0.2in}(2\log(y\msq^{2}-yz\mgluino^{2}+(z-xz)\mq^{2}-xyq^{2}
+x\lambda^{2})+
\nonumber\\
&&\hspace*{-0.2in}\left.\frac{(-2x-2y+xy+y^{2})\mgluino^{2}
+(1+x^{2}+xy)\mq^{2}
+(2x-xy)q^{2}+2z\mgluino\mq}{y\msq^{2}-yz\mgluino^{2}+
(z-xz)\mq^{2}-xyq^{2}
+x\lambda^{2}
})\right],\nonumber\\
&&\\
\label{susyF_{V2}}
\lefteqn{\hspace*{-0in}\tilde{F}_{\rm V2}=(C_{2}(N)-1/2C_{2}(G))
\feynint}\nonumber\\
&&\hspace*{-0.2in}\frac{2((1-y)\mgluino^{2}+(1-x)\mq^{2}
+(x+y)\mgluino\mq)}
{z\msq^{2}+(x-y+xy+y^{2})\mgluino^{2}+(-x+y+x^{2}+xy)\mq^{2}
-xyq^{2}},\\
\label{susyF_{V3}}
\lefteqn{\hspace*{-0in}\tilde{F}_{\rm V3}=C_{2}(G)
\left[\frac{1}{2\bar{\epsilon}}
+\frac{1}{2}\log
\mu^{2}-\frac{1}{4}-\feynint\right.}\nonumber\\
&&\hspace*{-0.2in}(\log(x\msq^{2}+(-y+z+xy+y^{2})\mgluino^{2}+(-x+x^{2}
+xy)\mq^{2}-xyq^
{2}+y\lambda^{2})+\nonumber\\
&&\hspace*{-0.2in}\left.\frac{(1+xy+y^{2})\mgluino^{2}
+(-2x-2y+x^{2}+xy)\mq^{2}
+y(2-x)q^{2}+2z\mgluino\mq}{2(x\msq^{2}+(-y+z+xy+y^{2})\mgluino^{2}
+(-x+x^{2}+xy)\mq^{2}
-xyq^{2}+y\lambda^{2})})\right],\nonumber\\
&&\\
\label{susyF_{V4}}
\lefteqn{\hspace*{-0in}\tilde{F}_{\rm V4}=C_{2}(G)
\left[\frac{2}{\bar{\epsilon}}
+2\log\mu^{2}-1-\feynint
\right.}\nonumber\\
&&\hspace*{-0.2in}(4\log((xy+y^{2})\mgluino^{2}+
(x^{2}+xy)\mq^{2}-xyq^{2}
+z\lambda^{2})
+\nonumber\\
&&\hspace*{-0.2in}\left.\frac{(1-x-2yz)\mgluino^{2}+(1-y-2xz)\mq^{2}
-(z+2xy)q^{2}-(x+y)\mgluino\mq}{(xy+y^{2})\mgluino^{2}
+(x^{2}+xy)\mq^{2}-xyq^{2}+z\lambda^{2}})\right].\nonumber\\
&&
\end{eqnarray}
Here $1/\bar{\epsilon}=1/\epsilon-\gamma_{\rm E}+\log4\pi$, $\mu$ is the
arbitrary renormalization scale, $x$, $y$, $z$ are the Feynman
parameters, $q^{2}$ is the external four-momentum squared for \gluon\
or \squark\  and all the other external particles (\quark\ and
\gluino) are on shell.  $C_{2}(R)$ is defined by
$t^{a}(R)t^{a}(R)=C_{2}(R)\cdot{\bf 1}$, where $t^{a}(R)$ is the
generator of SU(N) in the representation R with normalization
${\rm tr}[t^{a}(N)t^{b}(N)]=1/2\delta^{ab}\equiv{C}(N)
\delta^{ab}$ for the
fundamental representation $N$.

The one-loop corrections to the \gluino\ self energies are shown in 
Fig.~\ref{wgluino}, giving
\begin{eqnarray}
\label{F_gluino1}
\lefteqn{F_{\tilde{g}1}=C_{2}(G)\left[-\frac{1}{\bar{\epsilon}}
-\log\mu^{2}+\right.}\nonumber \\
&&2\wfeynint((1-x)\log(x\mgluino^{2}+(-x+x^{2})q^{2}
+y\lambda^{2})+\nonumber\\
&&\left.\frac{-4(-x+x^{2})\mgluino^{2}+2(1-x)(-x+x^{2})q^{2}}
{x\mgluino^{2}+(-x+x^{2})q^{2}+y\lambda^{2}})\right],\\
\label{F_gluino2}
\lefteqn{F_{\tilde{g}2}=C(N)\left[-\frac{2}{\bar{\epsilon}}-2\log\mu^{2}+
\right.}\nonumber \\
&&4\wfeynint((1-x)\log(y\msq^{2}+x\mq^{2}+(-x+x^{2})q^{2})+\nonumber\\
&&\left.\frac{2(1-x)(-x+x^{2})q^{2}}{y\msq^{2}+x\mq^{2}+(-x+x^{2})q^{2}})
\right]
.\hspace*{2in}
\end{eqnarray}
Here again the $q^{2}$ is the external four-momentum squared of the
\gluino, and is the external four-momentum squared of the 
\quark, \squark\ or \gluon\ in the \quark, \squark\ or \gluon\ 
self energies below.

With the same conventions we use above, we can get the corrections  
to the self
energies of \quark, \squark\ and \gluon: (the corresponding  
one-loop Feynman
diagrams are Fig.~\ref{wquark}, Fig.~\ref{wsquark}, and  
Fig.~\ref{wgluon})
\begin{eqnarray}
\label{F_quark1}
\lefteqn{F_{q1}=C_{2}(N)\left[-\frac{1}{\bar{\epsilon}}-\log\mu^{2}
+\right.}\nonumber \\
&&2\wfeynint(
(1-x)\log(x\mq^{2}+(-x+x^{2})q^{2}+y\lambda^{2})+\nonumber\\
&&\left.\frac{2(-x+x^{2})((1-x)q^{2}-2\mq^{2})}{x\mq^{2}
+(-x+x^{2})q^{2}
+y\lambda^{2}})\right],\hspace*{2in}\\
\label{F_quark2}
\lefteqn{F_{q2}=C_{2}(N)\left[-\frac{1}{\bar{\epsilon}}-
\log\mu^{2}+\right.}\nonumber\\
&&2\wfeynint(
x\log(x\msq^{2}+y\mgluino^{2}+(-x+x^{2})q^{2})+\nonumber\\
&&\left.\frac{2x(-x+x^{2})q^{2}}{x\msq^{2}+y\mgluino^{2}+
(-x+x^{2})q^{2}})\right],\\\label{F_squark1}
\lefteqn{F_{\tilde{q}1}=C_{2}(N)\left[-\frac{2}{\bar{\epsilon}}
-2\log\mu^{2}+
\frac{2}{3}-\right.}\nonumber\\
&&4\wfeynint(3(-x+x^{2})\log(x\mq^{2}+(-x+x^{2})q^{2}
+y\mgluino^{2})-\nonumber\\
&&\left.\frac{x(1-x)(-x+x^{2})q^{2}}{x\mq^{2}+(-x+x^{2})q^{2}
+y\mgluino^{2}})\right], \\
\label{F_squark2}
\lefteqn{F_{\tilde{q}2}=C_{2}(N)\left[\frac{2}{\bar{\epsilon}}+2\log\mu^{2}+
\frac{1}{6}-\right.}\nonumber\\
&&\wfeynint((4-6x+3x^{2})\log(x\msq^{2}+(-x+x^{2})q^{2}
+y\lambda^{2})+\nonumber\\
&&\left.\frac{(2-x)^{2}(-x+x^{2})q^{2}}{x\msq^{2}+(-x+x^{2})q^{2}
+y\lambda^{2}})\right],\\
\label{F_g1}
\lefteqn{F_{g1}=C_{2}(G)\left[\frac{5}{3\bar{\epsilon}}
+\frac{5}{3}\log\mu^{2}-\right.}\nonumber \\
&&\left.\wfeynint((1+4x-4x^{2})\log(\lambda^{2}+(-x+x^{2})q^{2})\right],\\
\label{F_g2}
\lefteqn{F_{g2}=C_{2}(G)\left[-\frac{2}{3\bar{\epsilon}}
-\frac{2}{3}\log\mu^{2}+\right.}\nonumber \\
&&\left.4\wfeynint(x(1-x)\log(\mgluino^{2}+(-x+x^{2})q^{2})\right],\\
\label{F_g3}
\lefteqn{F_{g3}=C(N)\left[-\frac{4}{3\bar{\epsilon}}
-\frac{4}{3}\log\mu^{2}+\right.}\nonumber \\
&&\left.8\wfeynint(x(1-x)\log(\mq^{2}+(-x+x^{2})q^{2})\right],\\
\label{F_g4}
\lefteqn{F_{g4}=C(N)\left[-\frac{2}{3\bar{\epsilon}}
-\frac{2}{3}\log\mu^{2}+\right.}\nonumber \\
&&\left.2\wfeynint((1-2x)^{2}\log(\msq^{2}+(-x+x^{2})q^{2})\right].
\end{eqnarray}
Notice that Fig.~\ref{wsquark} (3) does not contribute to the  
\squark\ self
energy corrections.  The expressions for $F_{g1}$ and $F_{g4}$ are
combinations of three or two diagrams.
\sect{The Super-Oblique Corrections}
\label{appB}
The super-oblique corrections comes from the gluon self energies  
$F_{g3}$,
$F_{g4}$ (Eq.~\ref{F_g3}, Eq.~\ref{F_g4}) and \gluino\ self energy
$F_{\tilde{g}2}$ (Eq.~\ref{F_gluino2}).  Then the relative difference of 
$g$ and $\tilde{g}$ is
\begin{equation}
\frac{g-\tilde{g}}{g}=
\frac{\alpha}{4\pi}F_{\rm ob}=\frac{\alpha}{4\pi}
\frac{F_{g3}+F_{g4}-F_{\tilde{g}2}}{2}.
\label{oblique}
\end{equation}
In the limit $\mq^{2}\ll{q}^{2}$ and $q^{2}\ll\msq^{2}$, these  
three integral
can easily be done and give
\begin{eqnarray}
F_{\tilde{g}2}&=&C(N)\left[-\frac{2}{\bar{\epsilon}}-2\log\mu^{2}
+2\log\msq^{2}-1\right],\\
F_{g3}&=&C(N)\left[-\frac{4}{3\bar{\epsilon}}-\frac{4}{3}
\log\mu^{2}+\frac{4}{3}\log{q^{2}}-\frac{20}{9}\right],\\
F_{g4}&=&C(N)\left[-\frac{2}{3\bar{\epsilon}}-\frac{2}{3}\log\mu^{2}
+\frac{2}{3}\log{\msq^{2}}\right].
\end{eqnarray}
Substituting them into Eq.~\ref{oblique} and taking $C(N)=1/2$, we have
\begin{equation}
\frac{g_{3}-\tilde{g}_{3}}{g_{3}}=\frac{\alpha_3}{4\pi}
\left(\frac{1}{3}\log\frac{q^{2}}{\msq^{2}}-\frac{11}{36}\right)
=\frac{\alpha_3}{12\pi}\left(\log\frac{q^{2
}}{\msq^{2}}-\frac{11}{12}\right).
\end{equation}
Here we use SU(3) as an example and sum over the contribution from
both $\tilde{q}_{L}$ and $\tilde{q}_{R}$.  If we want to separate
$\tilde{q}_{L}$ and $\tilde{q}_{R}$ (since their mass can be
different), then the result is half as big, which is just the
the formula given in the paper (Eq.~\ref{simpli_ob}) when 
$q^2 \sim m_{\tilde g}^2$.
The super-oblique corrections we give here take
into account the finite contribution even from the light  quark states. 
They are not included in the renormalization of the
coupling constant in an effective field
theory approach as was done in Ref.~\cite{feng}.  However, the
additional loop from light states generates a universal
finite term which one might as well include in the definition
of the oblique parameter, as it will in any case be included
for the physical predictions (assuming one is incorporating
the nonlogarithmically enhanced terms).
 
\sect{Gluon Emission For $\tilde{q}\rightarrow{q}\tilde{g}$}
\label{appC}
In order to cancel the infrared divergences (which come from the
massless gluon) in the vertex and self energy corrections, we need
to include $\tilde{q}\rightarrow{q}\tilde{g}g$ in calculating the
decay rate for $\tilde{q}\rightarrow{q}\tilde{g}$.   There are three
diagrams contributing to the gluon emission in the process
$\tilde{q}\rightarrow{q}\tilde{g}$ (Fig.~\ref{emission}).  The matrix
element $M=M_{1}+M_{2}+M_{3}$, where
\begin{eqnarray}
\label{M1}
M_{1}&\!\!\!\!=&\!\!\!\!-\frac{i\tilde{g}^{2}}{\sqrt{2}}\bar{u}_{q}(p',\mq)
\gamma_{\mu}t^{b}\epsilon^{*{b}}_{\mu}(k)
\frac{p\llap{/}'+k\llap{/}+\mq}{(p'+k)^{2}-\mq^{2}}(1+
\gamma_{5})t^{a}\nu_{\tilde{g}}(p,\mgluino),\\
\label{M2}
M_{2}&\!\!\!\!=&\!\!\!\!-\frac{i\tilde{g}^{2}}{\sqrt{2}}\bar{u}_{q}(p',\mq)
(1+\gamma_{5})t^{a}
\frac{(2q-k)^{\mu}}{(q-k)^{2}-\msq^{2}}t^{b} 
\epsilon^{*{b}}_{\mu}(k)\nu_{\tilde{g}}(p,\mgluino),\\
\label{M3}
M_{3}&\!\!\!\!=&\!\!\!\!\frac{\tilde{g}^{2}}{\sqrt{2}}f^{abc}\bar{u}_{q}
(p',\mq)(1+\gamma_{5})t^{c}\frac{p\llap{/}
+k\llap{/}-\mgluino}{(p+k)^{2}-\mgluino^{2}}\gamma^{\mu}
\epsilon^{*{b}}_{\mu}(k)\nu_{\tilde{g}}(p,\mgluino).
\end{eqnarray}
The decay rate for on shell \squark\ decay 
$\tilde{q}\rightarrow{q}\tilde{g}g$  is given by 
\begin{equation}
{\rm d}\Gamma=\frac{1}{(2\pi)^{3}}\frac{1}{32\msq^{3}}
\left|M\right|^{2}\,{\rm d}u\,{\rm d}s,  
\label{decayrate}
\end{equation}
where $\left|M\right|^{2}$ can be expressed in terms of $s=(p+k)^{2},
t=(p'+k)^{2}, u=(p+p')^2$, ($s+t+u=\msq^{2}+\mgluino^{2}+\mq^{2}$)
$\msq$, $\mgluino$, $\mq$ and $\lambda$.  Integrating over the whole
phase space, we can get the correction $F_{\rm emission}$, which is given
by
\begin{equation}
\Gamma_{\tilde{q}\rightarrow{q}\tilde{g}g}=
\frac{\tilde\alpha}{4\pi}\tilde\Gamma_{0}F_{\rm emission},
\label{defgamma}
\end{equation}
 where $\tilde\Gamma_{0}$ is the tree level decay rate for
$\tilde{q}\rightarrow{q}\tilde{g}$:
\begin{equation}
\tilde\Gamma_{0}=\frac{C_{2}(N)\tilde{g}^{2}}{2\pi{\msq}^{2}}\left|{\bf 
p}\right|\left[E_{q}E_{\tilde{g}}+\left|{\bf p}\right|^{2}\right].
\label{gamma0}
\end{equation}
Here $\left|{\bf p}\right|$, $E_{q}$, $E_{\tilde{g}}$ are all  
measured in the
rest frame of \squark.
\sect{$O(\alpha)$ Corrections To Couplings In the Exact SUSY Limit}
\label{appD}
In the SUSY limit, when $\mq=\msq$ and $\mgluino=0$, supersymmetry is
exactly preserved, so the supersymmetric Yukawa coupling 
$\tilde{g}_{\rm i}$ is
equal to the Standard Model gauge coupling $g_{\rm i}$ 
(${\rm i}=1,2,3$ indexing
U(1), SU(2) and SU(3) respectively).  Using the formulas given
in Appendix~A for the vertex and self energy diagrams, we show
explicitly in Table~\ref{SUSYlimit1} and \ref{SUSYlimit2}\footnote{\#  in
Table~\ref{SUSYlimit1} and \ref{SUSYlimit2} means any finite number coming from
calculation.} that to the first order, $g_{\rm i}=\tilde{g}_{\rm i}$.  The
vanishing of the $C_2(N)$ part is guaranteed by the Ward-Identity.
The coefficient of $\log\lambda^{2}$ is the same as $\log\mu^{2}$,
which  had to be true on dimensional grounds.  Notice that there are still 
infrared divergences
(the coefficient for $\log\lambda^{2}$ is not zero). 
Presumably these will cancel against emission diagrams
and other diagrams which don't appear away from the supersymmetric limit,
like  the collinear divergence coming from
$g\rightarrow\tilde{g}\tilde{g}$. 
\begin{table}
\begin{tabular}{ccccccc} \hline
&\multicolumn{3}{c}{$C_{2}(N)$}&\multicolumn{3}{c}
{$C_{2}(G)$} \\ \cline{2-7}
&$\log\mu^{2}$&$\log\lambda^{2}$&\#&\hspace*{0.2in}$
\log\mu^{2}$&$\log\lambda^{2}$&\#\hspace*{0.7in}\\ \hline
$\hspace*{0.7in}F_{v1}$&\hspace*{0.1in}1&2&5&\hspace*{0.2in}
-1/2&-1&-5/2\hspace*{0.7in}\\\hline
$\hspace*{0.7in}F_{v2}$&\hspace*{0.1in}1&0&3&\hspace*{0.2in}
-1/2&0&-3/2\hspace*{0.7in}\\\hline
$\hspace*{0.7in}F_{v3}$&\hspace*{0.1in}0&0&0&\hspace*{0.2in}
3/2&0&5/2\hspace*{0.7in}\\\hline
$\hspace*{0.7in}F_{v4}$&\hspace*{0.1in}0&0&0&\hspace*{0.2in}
1/2&0&3/2\hspace*{0.7in}\\\hline
$\hspace*{0.7in}F_{q1}$&\hspace*{0.1in}-1&-2&-5&\hspace*{0.2in}
0&0&0\hspace*{0.7in}\\\hline
$\hspace*{0.7in}F_{q2}$&\hspace*{0.1in}-1&0&-3&\hspace*{0.2in}
0&0&0\hspace*{0.7in}\\\hline
$\hspace*{0.7in}1/2F_{g1}$&\hspace*{0.1in}0&0&0&\hspace*{0.2in}
5/6&-5/6&0\hspace*{0.7in}\\\hline
$\hspace*{0.7in}1/2F_{g2}$&\hspace*{0.1in}0&0&0&\hspace*{0.2in}
-1/3&1/3&0\hspace*{0.7in}\\\hline
$\hspace*{0.7in}F_{\rm total}$&\hspace*{0.1in}0&0&0&\hspace*{0.2in}
3/2&-3/2&0\hspace*{0.7in}\\\hline
\end{tabular}
\caption{$O(\alpha)$ Correction to the gauge coupling constant $g$ in the  
SUSY limit }
\label{SUSYlimit1}
\end{table}

\begin{table}
\begin{tabular}{ccccccc} \hline
&\multicolumn{3}{c}{$C_{2}(N)$}&\multicolumn{3}{c}{$C_{2}(G)$} 
\\ \cline{2-7}
&$\log\mu^{2}$&$\log\lambda^{2}$&\#&$\log\mu^{2}$&$\log
\lambda^{2}$&\#\hspace*{0.7in}\\ \hline
$\hspace*{0.7in}\tilde{F}_{v1}\hspace*{0.1in}$&1&2&4&\hspace*{0.2in}
-1/2&-1&-2\hspace*{0.7in}\\\hline
$\hspace*{0.7in}\tilde{F}_{v2}\hspace*{0.1in}$&0&0&2&\hspace*{0.2in}
0&0&-1\hspace*{0.7in}\\\hline
$\hspace*{0.7in}\tilde{F}_{v3}\hspace*{0.1in}$&0&0&0&\hspace*{0.2in}1/2
&-1/2&0\hspace*{0.7in}\\\hline
$\hspace*{0.7in}\tilde{F}_{v4}\hspace*{0.1in}$&0&0&0&\hspace*{0.2in}2
&-1/2&3\hspace*{0.7in}\\\hline
$\hspace*{0.7in}1/2F_{q1}\hspace*{0.1in}$&-1/2&-1&-5/2&\hspace*{0.2in}0
&0&0\hspace*{0.7in}\\\hline
$\hspace*{0.7in}1/2F_{q2}\hspace*{0.1in}$&-1/2&0&-3/2&\hspace*{0.2in}
0&0&0\hspace*{0.7in}\\\hline
$\hspace*{0.7in}1/2F_{\tilde{q}1}\hspace*{0.1in}$&-1&0&-2&\hspace*{0.2in}
0&0&0\hspace*{0.7in}\\\hline
$\hspace*{0.7in}1/2F_{\tilde{q}2}\hspace*{0.1in}$&1&-1&0&\hspace*{0.2in}
0&0&0\hspace*{0.7in}\\\hline
$\hspace*{0.7in}1/2F_{\tilde{g}}\hspace*{0.1in}$&0&0&0&\hspace*{0.2in}
-1/2&1/2&0\hspace*{0.7in}\\\hline
$\hspace*{0.7in}\tilde{F}_{\rm total}$&0&0&0&\hspace*{0.2in}3/2&-3/2&
0\hspace*{0.7in}\\\hline
\end{tabular}
\caption{$O(\alpha)$ Correction to the Yukawa coupling constant 
$\tilde{g}$ in the  
SUSY limit}
\label{SUSYlimit2}
\end{table}

\sect{The Cancellation Of The Infrared Divergences}
\label{appE}
Some of the vertices and self energies corrections have infrared
divergences when the momentum goes to zero.  That is why we introduce
an infinitesimal gluon mass $\lambda$ as a regulator.  This infrared
divergences cannot be cancelled unless we add the contribution of
$\tilde{q}\rightarrow{q}\tilde{g}g$ to
$\tilde{q}\rightarrow{q}\tilde{g}$ process, so that the decay rate for
$\tilde{q}\rightarrow{q}\tilde{g}$\ is finite.  This is also necessary
because experimentally we cannot separate these two processes due to
the finite resolution of the detector.

First let us study the infrared divergence of the $M_{1}M_{1}^{*}$
($M_{1}$ is given in Eq.~\ref{M1}) in $|M|^{2}$ for
$\tilde{q}\rightarrow{q}\tilde{g}g$.  When the gluon momentum $k$ is
very small, we can find the infrared divergences by keeping only the
leading order term in $k$ in both the numerator and denominator.
Under this approximation, we can treat gluon as massless until we do
the final integral over $k_{0}$, which would be divergent in the lower
limit if the gluon is really massless.  We then let the gluon have an
infinitesinal mass $\lambda$, which would then provide a lower cut off
for the integral.  For a massless gluon, $k=(k_{0}, \vec{{\bf k}})$
with $k_{0}=|\vec{{\bf k}}|$.  We get
\begin{eqnarray}
M_{1}M_{1}^{*}&\approx&{-}4C_{2}(N)\tilde{g}^{2}p\cdot{p}'
\frac{4C_{2}(N)\tilde{g}^{2}\mq^{2}}{
\left[(p'+k)^{2}-\mq^{2}\right]^{2}}\nonumber \\
&=&-4C_{2}(N)\tilde{g}^{2}p\cdot{p}'\frac{C_{2}(N)\tilde{g}^{2}
\mq^{2}}{(p'\cdot{k})^{2}}.
\end{eqnarray}
The $4C_{2}(N)\tilde{g}^{2}p\cdot{p}'$ is the matrix element squared for the
tree level $\tilde{q}\rightarrow{q}\tilde{g}$, which will give
$\tilde\Gamma_{0}$ after we perfom the phase space integral of \quark\ and
\gluino.  According to Eq.~\ref{defgamma}, we have
\footnote{$F_{\rm emission}^{\rm ij}$ is the contribution of the
$F_{\rm emission}$ from $M_{\rm i}M_{\rm j}^{*}$.}
\begin{eqnarray}
F_{\rm emission}^{11}&=&-C_{2}(N)(4\pi)^{2}\kint
\frac{\mq^{2}}{(p'\cdot{k})^
{2}}\nonumber \\
&=&-C_{2}(N)(4\pi)^{2}\int_{\lambda}\frac{k_{0}^{2}
{\rm d}k_{0}}{4\pi^{2}k_{0}}\int
\frac{{\rm d} \Omega}{4\pi}\frac{\mq^{2}}{(p'\cdot{k})^{2}}\nonumber\\
&=&-C_{2}(N)(4\pi)^{2}\int_{\lambda}\frac{k_{0}^{2}{\rm d}
k_{0}}{4\pi^{2}k_{0}^{3}}=2C_{2}(N)\log\lambda^{2}.\hspace*{1in}
\end{eqnarray}
Similarly we have
\begin{eqnarray}
F_{\rm emission}^{22}&=&2C_{2}(N)\log\lambda^{2},\\
F_{\rm emission}^{33}&=&2C_{2}(G)\log\lambda^{2}.
\end{eqnarray}
The analysis of the cross term $M_{\rm i}M_{\rm j}^{*}$ is a little bit more 
complicated, but similar.  Considering $F_{\rm emission}^{12}$ as an  
example, we
get\footnote{$\hat{k}=k/k_{0}$}
\begin{eqnarray}
\lefteqn{\hspace*{-0in}F_{\rm emission}^{12}=(C_{2}(N)-
\frac{1}{2}C_{2}(G))(4\pi)^{2}
\kint\frac{2p'\cdot{q}}{(p'\cdot{k})(q\cdot{k})}}\nonumber\\
&&\hspace*{-0.22in}=(C_{2}(N)-\frac{1}{2}C_{2}(G))(4\pi)^{2}\int_{\lambda}
\frac{k_{0}^{2}{\rm d}k_{0}}
{4\pi^{2}k_{0}^{3}}\int\frac{d\Omega}{4\pi}\frac{\msq^{2}
+\mq^{2}-\mgluino^{2}}
{(p'\cdot{\hat{k}})(q\cdot{\hat{k}})}\nonumber\\
&&\hspace*{-0.22in}=(C_{2}(N)-\frac{1}{2}C_{2}(G))(4\pi)^{2}\int_{\lambda}
\frac{(\msq^{2}+\mq^{2
}-\mgluino^{2}){\rm d}k_{0}}{4\pi^{2}k_{0}}\int^{1}_{0}
{\rm d}\xi\frac{1}{(\xi{p'}+(1-\xi)q
)^{2}}\nonumber\\
&&\hspace*{-0.22in}=-2(C_{2}(N)-\frac{1}{2}C_{2}(G))\log\lambda^{2}\int^{1}_{0}
{\rm d}\xi\frac{\msq^{2
}+\mq^{2}-\mgluino^{2}}{\xi^{2}\mgluino^{2}+\xi(\mq^{2}-\msq^{2}
-\mgluino^{2})+\msq^{2}},\nonumber\\
&&
\end{eqnarray}
where to get from step 2 to step 3, we introduced the Feynman  
parameter $\xi$
and performed the ${\rm d}\Omega$ integral.  In the last step, we keep
 only the term
proportional to $\log\lambda^{2}$ and 
neglecting all the other terms.  Similarly, we get
\begin{eqnarray}
\lefteqn{\hspace*{-0in}F_{\rm emission}^{13}=}\nonumber\\
&&\hspace*{-0.25in}-C_{2}(G)\log\lambda^{2}\int^{1}_{0}{\rm d}
\xi\frac{\msq^{2}-\mq^{2
}-\mgluino^{2}}{\xi^{2}(2\mgluino^{2}+2\mq^{2}-\msq^{2})+
\xi(\msq^{2}-3\mq^{2}-
\mgluino^{2})+\mq^{2}},\nonumber\\
&&
\end{eqnarray}
\begin{equation}
F_{\rm emission}^{23}=-C_{2}(G)\log\lambda^{2}\int^{1}_{0}{\rm d}
\xi\frac{\msq^{2}+
\mgluino^{2}-\mq^{2}}{\xi^{2}\mq^{2}+\xi(\mgluino^{2}-m_q^{2}
-\msq^{2})+\msq^{2}}.
\end{equation}

Now that we have calculated the infrared divergences in the gluon emission,
let us study the infrared divergences of the {\it \~{q}q\~{g}} vertex
corrections when \squark\ is on shell.  Only $\tilde{F}_{\rm V1}$,
$\tilde{F}_{\rm V3}$, $\tilde{F}_{\rm V4}$ have infrared divergences.  For
$\tilde{F}_{\rm V1}$ (Eq.~\ref{susyF_{V1}}), the denominator of the second
integral is divergent only when $x\rightarrow1$ and
$y,z\rightarrow0$. In this region, we can set $x=1$ and $y=z=0$ in the
numerator of Eq.~\ref{susyF_{V1}}.  We can also set $x=1$ in the
$\lambda^{2}$ term in the denominator.  Keeping only the term
proportional to $\log\lambda^{2}$ and neglecting all the other terms,
we have\cite{peskin}
\begin{eqnarray}
\lefteqn{\hspace*{-0in}\tilde{F}_{\rm V1}=(C_{2}(N)-1/2C_{2}(G))}\nonumber\\
&&\hspace*{-0in}\int{{\rm d}x}{\rm d}y
\frac{2(\msq^{2}+\mq^{2}-
\mgluino^{2})}{(y-xy)\msq^{2}-y(1-x-y)\mgluino^{2}+(1-x)(1-x-y)\mq^{2}
+\lambda^{2}}.\nonumber\\
&&
\end{eqnarray}
Let $y=(1-x)\xi$, $\omega=(1-x)$, this expression becomes
\begin{eqnarray}
\lefteqn{\hspace*{-0in}\tilde{F}_{\rm V1}=-(C_{2}(N)-1/2C_{2}(G))}\nonumber\\
&&\hspace*{-0in}\int^{1}_{0}{{\rm d}\xi}
\int^{1}_{0}{\rm d}{\omega}
^{2}\frac{\msq^{2}+\mq^{2}-\mgluino^{2}}{\omega^{2}
\left[\xi\msq^{2}-\xi(1-\xi)
\mgluino^{2}+(1-\xi)\mq^{2}\right]+\lambda^{2}}.
\end{eqnarray}
Performing the $\omega^{2}$ integral and let  
$\xi\rightarrow{1-\xi}$, the
result is
\begin{eqnarray}
\tilde{F}_{\rm V1}&\!\!\!\!=\!\!\!\!&(C_{2}(N)-\frac{1}{2}C_{2}(G))\log
\lambda^{2}\int^{1}_{0}{\rm d}\xi
\frac{\msq^{2}+\mq^{2}-\mgluino^{2}}{\xi^{2}\mgluino^{2}+\xi(\mq^{2}-\msq^{2}
-\mgluino^{2})+\msq^{2}},\nonumber\\
&&
\end{eqnarray}
which is exactly the same as $-F_{\rm emission}^{12}/2$.

Performing the same calculation, we see that the infrared divergences
of $2\tilde{F}_{\rm V3}$ and $2\tilde{F}_{\rm V4}$ cancel 
$F_{\rm emission}^{23}$
and $F_{\rm emission}^{13}$, and the infrared divergences of $F_{q1}$,
$F_{\tilde{q}2}$ and $F_{\tilde{g}}$ cancel $F_{\rm emission}^{11}$,
$F_{\rm emission}^{22}$, $F_{\rm emission}^{33}$. Thus we have shown explicitly that
the infrared divergences of the vertices, self energies and gluon
emission do cancel with each other and give an infrared convergent
result at the end.

\sect{The Super-Oblique Corrections From a Messenger Sector}
\label{appF}
In this section, We consider the super-oblique contribution from a sample
messenger sector which is in the 5, $\overline{5}$ or 10,
$\overline{10}$ representation of  $SU(5)$. We use $m_{\rm f}$, 
$m_{\rm s}$ to
stand for the mass of  the fermion and the scalar, where 
$m_{\rm f}=\lambda S$ and
$m_{\rm s}=\sqrt{\lambda^2 S^2\pm\lambda{F}_{\rm S}}$.  In addition, we define
$m_{\rm s}^2/m_{\rm f}^2=1\pm{x}$, thus 
$x=\lambda{F_{\rm S}}/\lambda^2S^2$.  In
the limit $q^2\ll{m}_{\rm s}^2$, and $q^2\ll{m}_{\rm f}^2$, 
Eq.~\ref{F_gluino2},
Eq.~\ref{F_g3} and Eq.~\ref{F_g4} give
\begin{eqnarray}
F_{\tilde{g}2}&=&C(N)\left[-\frac{2}{\bar{\epsilon}}-2\log\mu^{2}
+2\log{m}_{\rm s}^{2}\right.\nonumber\\
&&\left.+\frac{-m_{\rm s}^4+4m_{\rm s}^2m_{\rm f}^2-3m_{\rm f}^4
-2m_{\rm f}^4\log(m_{\rm s}^2/m_{\rm f}^2)}{(-m_{\rm s}^2+m_{\rm f}^2)^2}
\right],\\
F_{g3}&=&C(N)\left[-\frac{4}{3\bar{\epsilon}}-\frac{4}{3}\log\mu^{2}
+\frac{4}{3}\log{m_{\rm f}^{2}}\right],\\
F_{g4}&=&C(N)\left[-\frac{2}{3\bar{\epsilon}}-\frac{2}{3}\log\mu^{2}
+\frac{2}{3}\log{m_{\rm s}^{2}}\right].
\end{eqnarray}
Substituting them into Eq.~\ref{oblique} and multiplying by another 1/2 factor as 
we explained in Appendix \ref{appB}, we get
\begin{equation}
\frac{g_{\rm i}-\tilde{g}_{\rm i}}{g_{\rm i}}=
\frac{\alpha}{24\pi}\left[-\frac{1}{2}\log(1-x^2)
+\frac{3(x^2+\log(1-x^2))}{4x^2}\right].
\label{messeq}
\end{equation}
The correction is large only when $x$ is very close to 1.  
In the limit that 
$|x|\ll{1}$, the correction is
\begin{equation}
\frac{g_{\rm i}-\tilde{g}_{\rm i}}{g_{\rm i}} =\frac{\alpha_{\rm i}}{192\pi}x^2.
\end{equation}
Notice that for U(1) coupling, we need to multiply by the hypercharge 
squared.  Also, there is an additional factor of 2 for  the U(1) 
since there is no $C(N)=1/2$ multiplicity factor.
If the messenger sector is in the 5, $\overline{5}$ or 10, 
$\overline{10}$ of the SU(5) representation, we can decompose it under 
${\rm SU(3)}\times{\rm {SU}(2)}_{\rm Y}$
\begin{eqnarray}
5&=&(3,1)_{-1/3}+(1,2)_{1/2},\\
10&=&(3,2)_{1/6}+(\overline{3},1)_{-2/3}+(1,1)_{1}.
\end{eqnarray}
Each SU(3) triplet or SU(2) doublet will contribute to the
super-oblique corrections.  In 5, there is one SU(3) triplet and one
SU(2) doublet, while for 10, there are three SU(3) triplets and
three SU(2) doublets.  Thus the multiplicity factor is $2N$, where
$N=(n_5+3n_{10})$ if we have $n_5$ pairs of 5 and $\overline{5}$,
$n_{10}$ pairs of 10 and $\overline{10}$.  The super-oblique
correction from the messenger sector is then
\begin{equation}
\frac{g_{\rm i}-\tilde{g}_{\rm i}}{g_{\rm i}}=\frac{\alpha_{\rm i}}{96\pi}Nx^2.
\label{nonstan}
\end{equation}
Substituting the actual numbers into Eq.~\ref{nonstan}, we get 
\begin{eqnarray}
SU(3)\ \ \ \ \ \ \frac{g_{3}-\tilde{g}_3}{g_3}=\ 0.04\%Nx^2,\\
SU(2)\ \ \ \ \ \ \frac{g_{2}-\tilde{g}_2}{g_2}=0.011\%Nx^2,\\
U(1)\ \ \ \ \ \frac{g_{1}-\tilde{g}_1}{g_1}=0.0056\%Nx^2,
\end{eqnarray}   
which is   very small.

\newpage

\large
\noindent
{\bf Figure Captions}
\vskip.5in
\normalsize

\begin{description}

\item[Figure 1.] One-loop diagrams for the {\it gqq} vertex.  The arrow  
shows the glow
of quark number.  \squark\ can be either $\tilde{q}_{\rm L}$ or  
$\tilde{q}_{\rm R}$.
$F_{\rm Vi}$ corresponds to the correction from diagram (i),  
${\rm i}=1,2,3,4$.  Similar
notations are used in the diagrams below.

\item[Figure 2.] One-loop diagrams for the {\it q\~{q}\~{g}} vertex.

\item[Figure 3.] One-loop diagrams for \gluino\ self energies.

\item[Figure 4.] One-loop diagrams for \quark\ self energies.
\item[Figure 5.] One-loop diagrams for \squark\ self energies.
\item[Figure 6.] One-loop diagrams for \gluon\ self energies.
\item[Figure 7.] Feynman diagrams for $\tilde{q}\rightarrow{q}\tilde{g}g$.
\item[Figure 8.] Number of events per year for  
$e^+e^-\rightarrow{t}\,\tilde{t}\,\tilde{g}$ at
$E_{\rm cm}=1000$ GeV and luminosity $L=5\times{10}^{33} {\rm cm}^{-2}{\rm s}^{-1}$.
\item[Figure 9.] Number of events per year for  
$e^+e^-\rightarrow{b}\,\tilde{b}\,\tilde{g}$ at
$E_{\rm cm}=1000$ GeV and luminosity $L=5\times{10}^{33} {\rm cm}^{-2}{\rm s}^{-1}$.
\item[Figure 10.] SU(3) non-oblique corrections for on shell squark decay
$\tilde{q}\rightarrow{q}\tilde{g}$.
\end{description}

\newpage
\setcounter{figure}{0}
\renewcommand{\thefigure}{\arabic{figure}}
\begin{figure}
\PSbox{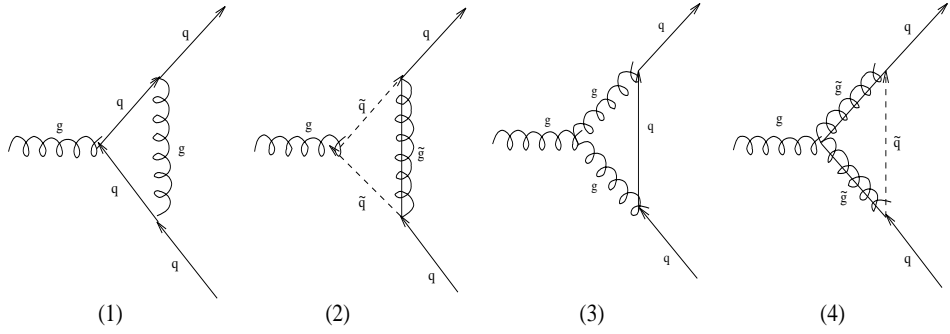 hoffset=-18 voffset=-256 hscale=70  
vscale=80}{4.8in}{1.4in}
\caption{One-loop diagrams for the {\it gqq} vertex.  The arrow  
shows the glow
of quark number.  \squark\ can be either $\tilde{q}_{\rm L}$ or  
$\tilde{q}_{\rm R}$.
$F_{\rm Vi}$ corresponds to the correction from diagram (i),  
${\rm i}=1,2,3,4$.  Similar
notations are used in the diagrams below.}
\label{smv}
\end{figure}
\begin{figure}
\PSbox{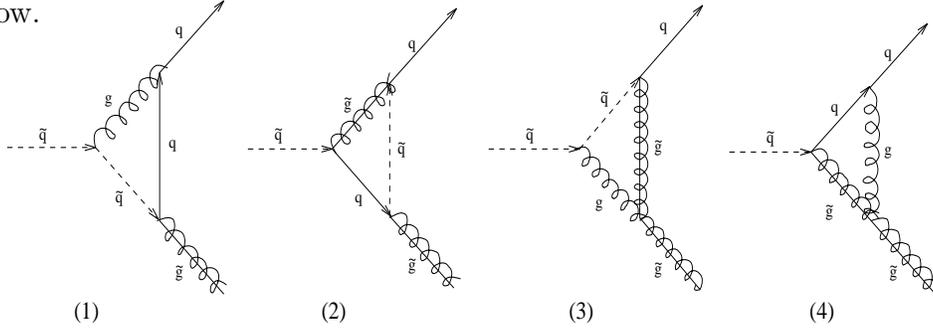 hoffset=-18 voffset=-256 hscale=70  
vscale=80}{4.8in}{1.4in}
\caption{One-loop diagrams for the {\it q\~{q}\~{g}} vertex.}
\label{susyv}
\end{figure}
\begin{figure}
\PSbox{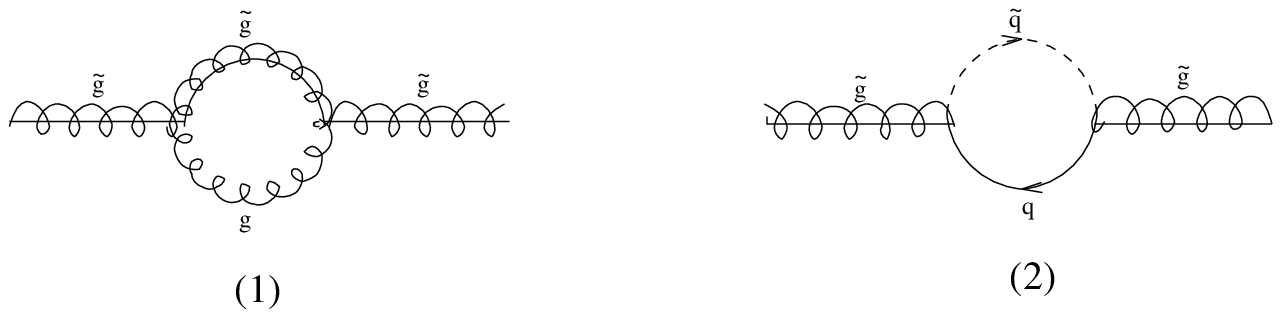 hoffset=-60 voffset=-280 hscale=80 
vscale=80}{4.8in}{1.0in}
\caption{One-loop diagrams for \gluino\ self energies.}
\label{wgluino}
\end{figure}
\begin{figure}
\PSbox{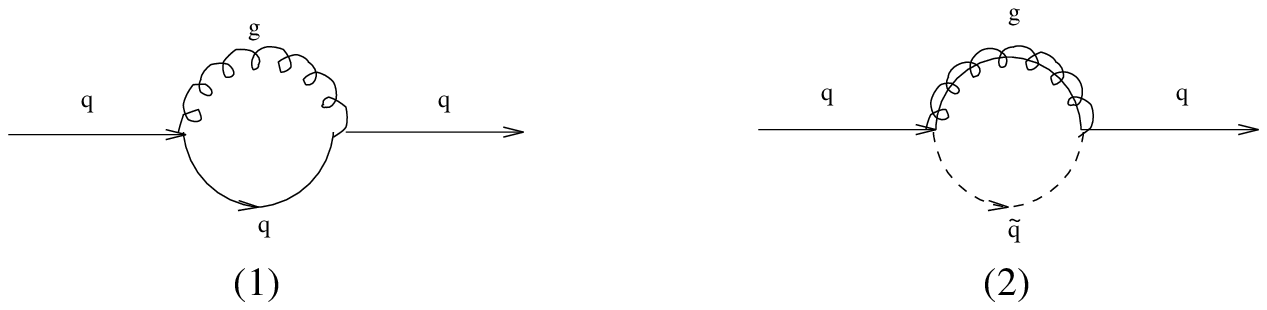 hoffset=-60 voffset=-280 hscale=80  
vscale=80}{4.8in}{1.0in}
\caption{One-loop diagrams for \quark\ self energies.}
\label{wquark}
\end{figure}
\begin{figure}
\PSbox{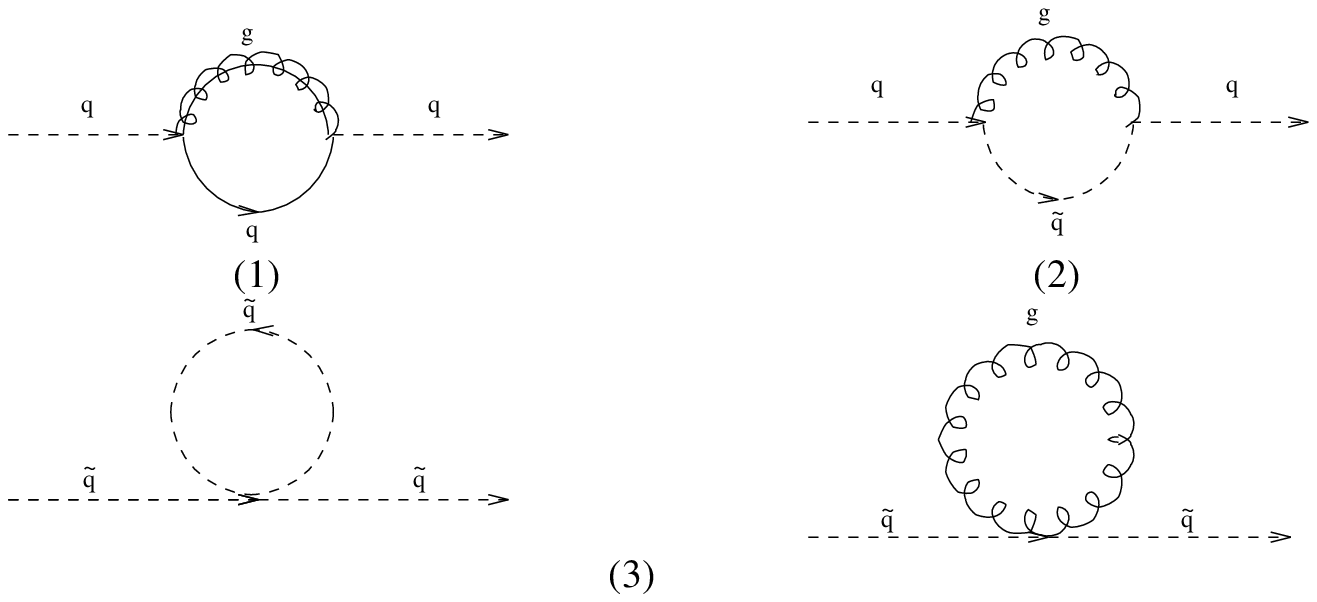 hoffset=-50 voffset=-250 hscale=80  
vscale=80}{4.8in}{1.4in}
\caption{One-loop diagrams for \squark\ self energies.}
\label{wsquark}
\end{figure}
\begin{figure}
\PSbox{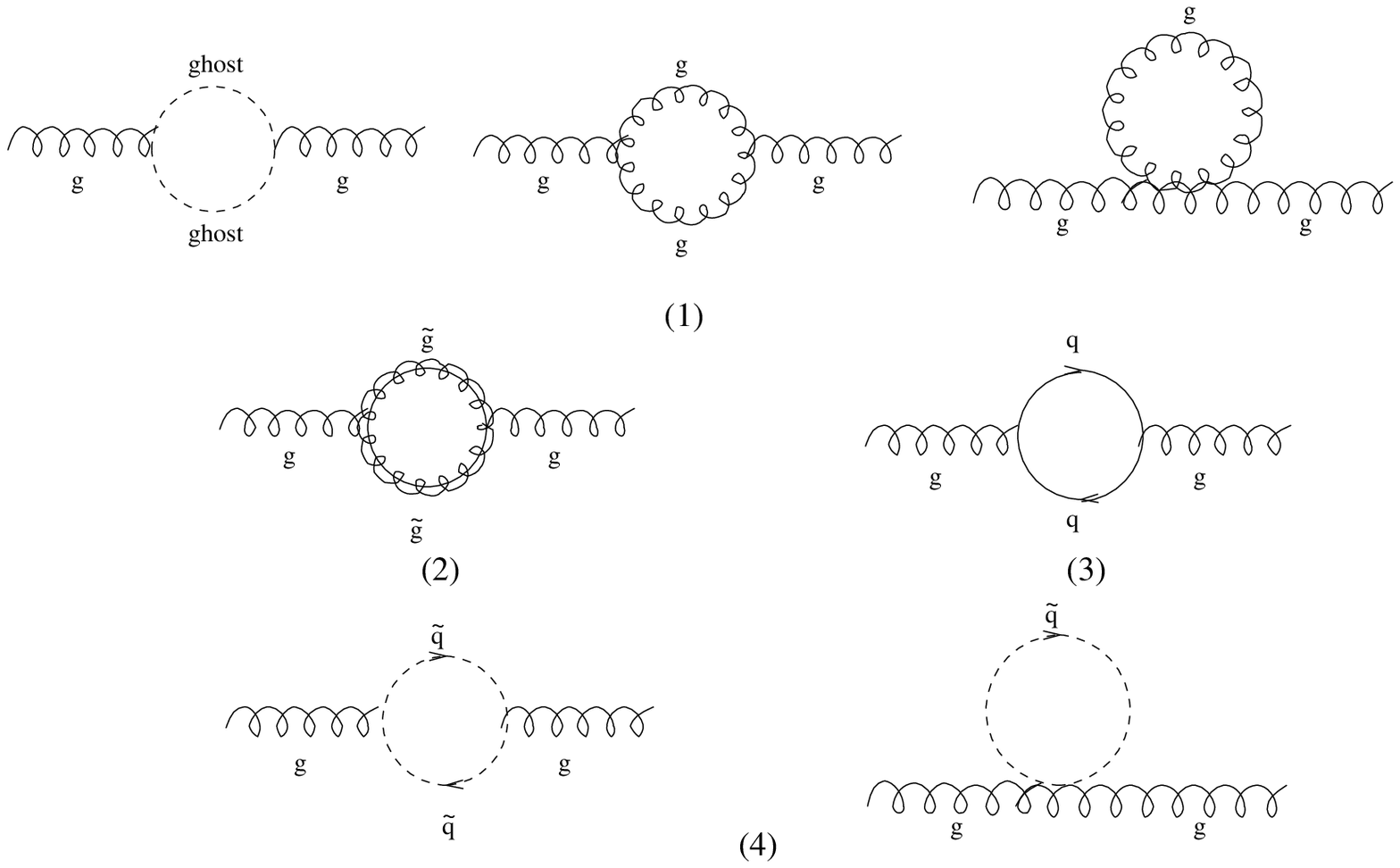 hoffset=-36 voffset=-200 hscale=80  
vscale=80}{4.8in}{2.8in}
\caption{One-loop diagrams for \gluon\ self energies.}
\label{wgluon}
\end{figure}
\begin{figure}
\PSbox{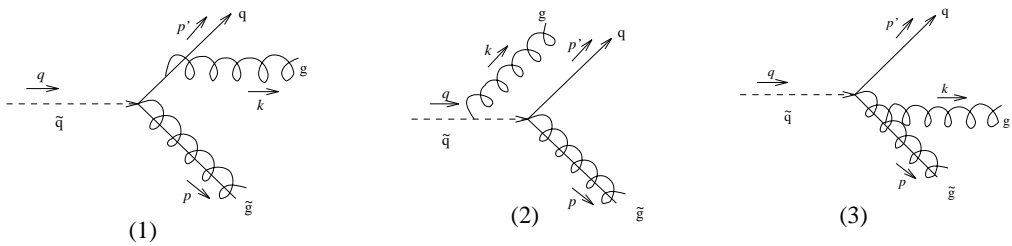 hoffset=-42 voffset=-280 hscale=80 
vscale=80}{4.8in}{1.1in}
\caption{Feynman diagrams for $\tilde{q}\rightarrow{q}\tilde{g}g$.}
\label{emission}
\end{figure}
\begin{figure}
\PSbox{ 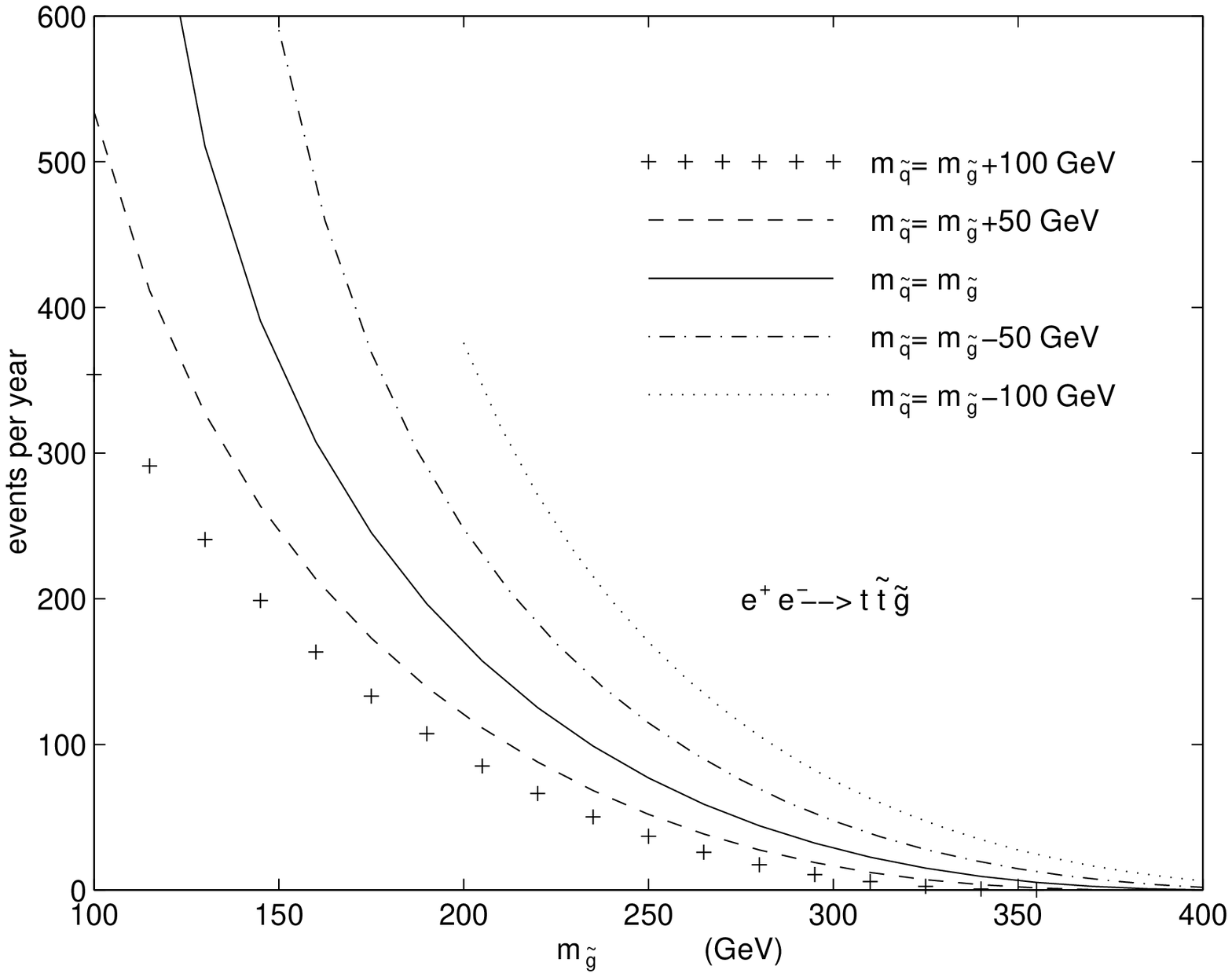 hoffset=36 voffset=-100 hscale=50 vscale=50}{4.8in}{3.0in}
\caption{Number of events per year for  
$e^+e^-\rightarrow{t}\,\tilde{t}\,\tilde{g}$ at
$E_{\rm cm}=1000$ GeV and luminosity $L=5\times{10}^{33} {\rm cm}^{-2}{\rm s}^{-1}$.}
\label{eet}
\end{figure}
\begin{figure}
\PSbox{ 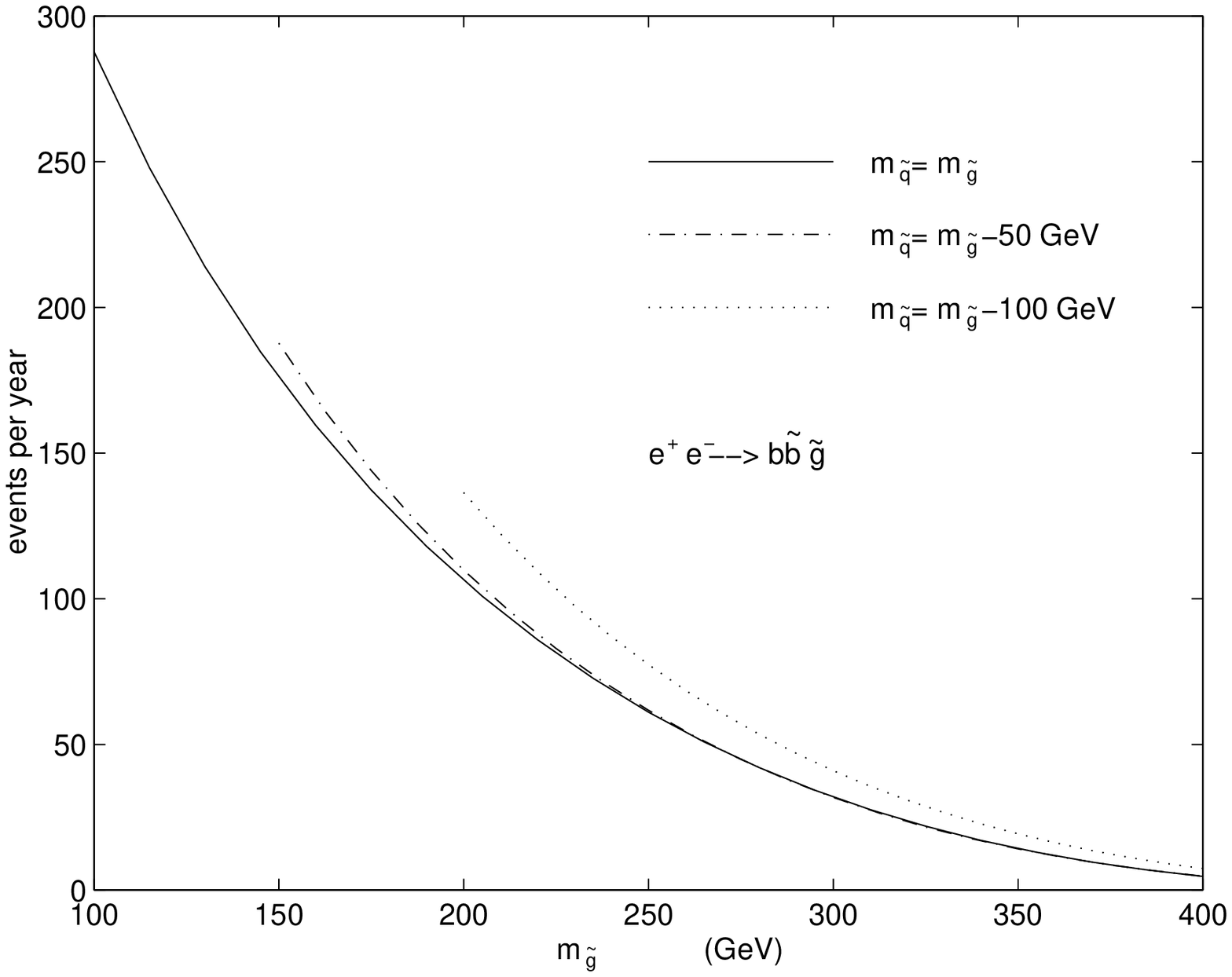 hoffset=36 voffset=-100 hscale=50 vscale=50}{4.8in}{3.0in}
\caption{Number of events per year for  
$e^+e^-\rightarrow{b}\,\tilde{b}\,\tilde{g}$ at
$E_{\rm cm}=1000$ GeV and luminosity $L=5\times{10}^{33} {\rm cm}^{-2}{\rm s}^{-1}$.}
\label{eeb}
\end{figure}
\begin{figure}
\PSbox{ 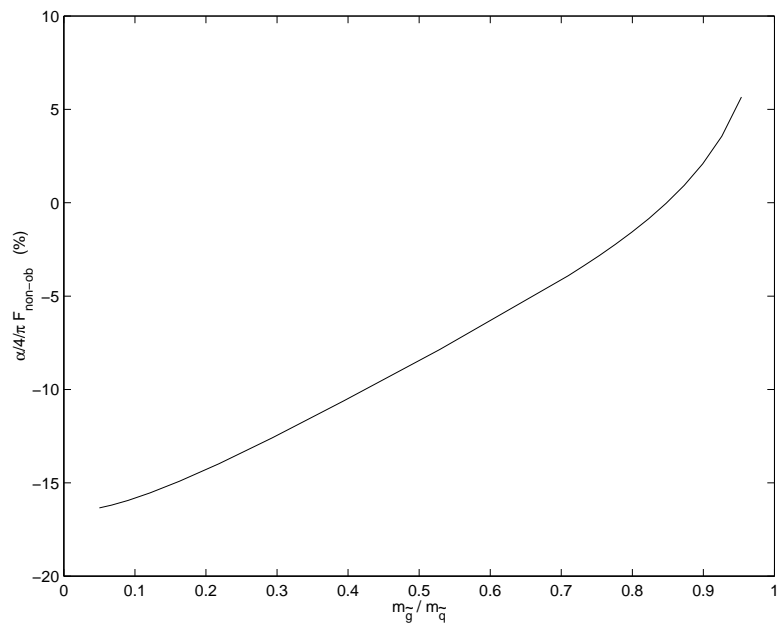 hoffset=36 voffset=-100 hscale=60  
vscale=60}{4.8in}{3.5in}
\caption{SU(3) non-oblique corrections for on shell squark decay
$\tilde{q}\rightarrow{q}\tilde{g}$.}
\label{changegluino}
\end{figure}

\end{document}